\documentclass[prb,superscriptaddress,twocolumn,showpacs]{revtex4}

\usepackage{amsmath,amssymb,graphicx}

\begin{document}

\title{Phase-field crystal study of grain-boundary premelting}

\author{Jesper Mellenthin}
\affiliation{Physique de la Mati\`ere Condens\'ee, Ecole Polytechnique, 
CNRS, 91128 Palaiseau, France}
\affiliation{Department of Physics, Northeastern University, 
Boston, Massachusetts 02115}
\author{Alain Karma}
\affiliation{Physics Department, Northeastern University, 
Boston, Massachusetts 02115}
\author{Mathis Plapp}
\affiliation{Physique de la Mati\`ere Condens\'ee, Ecole Polytechnique, 
CNRS, 91128 Palaiseau, France}

\begin{abstract}
We study the phenomenon of grain-boundary premelting for 
temperatures below the melting point in the phase-field crystal 
model of a pure material with hexagonal ordering in two 
dimensions. We investigate the structures of symmetric tilt 
boundaries as a function of misorientation $\theta$ for two 
different inclinations and compute in the grand canonical 
ensemble the ``disjoining potential" $V(w)$ that describes the 
fundamental interaction between crystal-melt interfaces as a 
function of the premelted layer width $w$, which is defined 
here in terms of the excess mass of the grain boundary 
via a Gibbs construction. The results reveal qualitatively different 
behaviors for high-angle grain boundaries that are uniformly wetted, with
$w$ diverging logarithmically as the melting point is approached from below, 
and low-angle boundaries that are punctuated by liquid pools surrounding 
dislocations, separated by solid bridges. The latter persist over a 
superheated range of temperature. This qualitative difference between high- 
and low- angle boundaries is reflected in the $w$-dependence of the disjoining 
potential that is purely repulsive ($V'(w)<0$ for all $w$) for
misorientations larger than a critical angle $\theta_c$, 
but switches from repulsive at small $w$ to attractive 
at large $w$ for $\theta<\theta_c$. In the latter case, $V(w)$ has a 
minimum that corresponds to a premelted boundary of finite width at 
the melting point. Furthermore, we find that the standard wetting 
condition $\gamma_{\mathrm gb}(\theta_c)=2\gamma_{\mathrm sl}$ gives a much 
too low estimate of $\theta_c$ when a low-temperature value of the grain
boundary energy $\gamma_{\mathrm gb}$ is used. In contrast, a reasonable 
lower-bound estimate can be obtained if $\gamma_{\mathrm gb}$ is 
extrapolated to the melting point, taking into account both the elastic 
softening of the material at high homologous temperature 
and local melting around dislocations.

\end{abstract}


\date{\today}

\maketitle

\section{Introduction and summary}
\label{sec:intro}
The presence of liquid films at grain boundaries   
for temperatures below the melting point can  
alter macroscopic properties of polycrystalline
materials and dramatically reduce resistance to shear stresses. The latter can 
lead to catastrophic material failure 
as exemplified by hot cracking during high-temperature processing of metallic
alloys \cite{Rappaz03,Wangetal04}.
While there is indirect experimental evidence for the occurrence
of grain-boundary premelting in both pure materials \cite{Glicksman72,Hsieh89} 
and alloys \cite{Wangetal04,Divinskietal05}, it is inherently difficult to 
image and to measure thermodynamic properties of
nanometer-width liquid films. One exception is optical microscopy of
colloidal crystals, which has produced striking ``atomistic''-scale
images of premelted grain boundaries \cite{Alsayed05}. Even in this case, however, 
the lack of precise control of grain geometry and external conditions makes 
it hard to determine the fundamental nature of the premelting transition.

Molecular dynamics (MD) simulations provide in principle 
a powerful alternative to experiment for studying grain-boundary
premelting. MD studies using Lennard-Jones \cite{Broughton86,Broughton87,Broughton98} and 
interatomic potentials for  
metals \cite{Nguyen92,Zhaoetal01,Fan05} and semiconductors such 
as silicon \cite{Alfetal07}
have reported evidence for disordered layers 
at grain boundaries at different temperatures
below \cite{Broughton86,Broughton87,Nguyen92,Broughton98,Zhaoetal01,Alfetal07} and
above \cite{Fan05} the melting point. In addition, 
such layers have been reported to exhibit fluid-like properties in a MD study of
grain-boundary shearing in a Lennard-Jones system where
the shear modulus decreased sharply below the melting point \cite{Broughton98}.
The large fluctuations inherent in MD simulations,
however, make it generally hard to compute precisely the thermodynamic
properties of grain boundaries at high homologous temperature and to
quantify the interaction between crystal-melt interfaces.

A remarkable study of grain-boundary premelting was carried
out by Kikuchi and Cahn \cite{Kikuchi80} using a lattice
gas model and a cluster variation approximation for the 
evaluation of its thermodynamic properties. Their results
were later corroborated by Monte Carlo simulations of the
same model \cite{Besold94}. They indeed found 
a liquid-like layer at the grain boundary for temperatures
well below the melting point. The width of this layer
diverges logarithmically when the melting point is approached.
While this study gave valuable insights, it did not yield
a complete picture of grain-boundary premelting since the 
construction of the lattice-gas model leads to numerous 
geometrical constraints, such that only a single 
misorientation could be investigated.

From a basic thermodynamic viewpoint, 
grain-boundary premelting is governed 
by the balance between bulk and interfacial free energies. 
While the difference in bulk free energies per unit volume
between solid and liquid \cite{remark1}, 
$\Delta G(T)\equiv G_s(T)-G_l(T)$, always favors a crystalline
state below the melting point, the interfacial free energy  
favors the formation of a liquid layer for wetting conditions. 
The total excess free energy (per unit area of grain boundary)
that reflects both contributions can be written in the 
form \cite{Smith53}
\begin{equation}
G_{\mathrm{exc}}(w,T)=\Delta G(T) w + 2\gamma_{\mathrm sl} + V(w),\label{Gw}
\end{equation}
where $w$ denotes the liquid layer width and
the last two terms on the right-hand-side
represent the interfacial free energy. The latter must reduce 
to twice the excess free energy of the solid-liquid
interface, $2\gamma_{\mathrm sl}$, when the two solid-liquid interfaces
are well separated, but generally contains an additional
contribution $V(w)$ when their separation 
becomes comparable to the intrinsic nanometer-width $\delta$
of an isolated solid-liquid interface. This additional
contribution, referred to hereafter as the ``disjoining
potential,'' represents the interaction due to the
overlap of two solid-liquid interfaces, which 
drives the formation of a liquid layer under wetting
conditions, or conversely joins two crystals for non-wetting
conditions. Its derivative $V'(w)$ is directly analogous to 
the disjoining pressure used in the physics of thin liquid 
films \cite{deGennes85}.

In this paper, we study grain-boundary premelting
using the phase-field crystal (PFC) modeling approach
\cite{Eldetal02,Eldetal04,Goletal05,Steetal06,Beretal06,Eldetal07}
inspired from classical density-functional theory \cite{RamYus79}.
In the present context, this mean-field approach
has the advantage of resolving the atomic-scale density-wave structure of  
a polycrystalline material while, at the same time, 
averaging out fluctuations. Therefore, it is ideally suited for computing 
quantitatively the disjoining potential and for elucidating its relationship 
to atomic grain-boundary structure. In a recent study, Berry {\it et al.}  
observed melting at grain boundaries in three-dimensional 
phase-field crystal simulations for bcc ordering \cite{Beretal08},
thereby suggesting the usefulness of this method for investigating 
fundamental aspects of this phenomenon. Thermodynamic
properties of premelted grain boundaries, however, were not studied
in detail in this work. The present work focuses on the
quantitative study of premelting in the general
framework of Eq.~(\ref{Gw}) with the appropriate choice of
thermodynamic ensemble for the phase-field crystal model, for
two-dimensional crystals with hexagonal symmetry.
We compute explicitly the dependence of the
disjoining potential on layer width and determine
wetting conditions as a function of grain-boundary orientation. This allows
us to make contact with sharp- \cite{Broughton87,Rappaz03}
and diffuse-interface \cite{Rappaz03,Lobkowski02,Tang06} theories of 
interfacial premelting. Like MD 
studies with truncated short-range interatomic
potentials \cite{Broughton86,Broughton87,Nguyen92,Broughton98,Zhaoetal01,Fan05,Alfetal07}, 
these theories neglect the effects of long-range dispersion forces  
considered in statistical theories of grain-boundary melting \cite{Lipowsky86}
and in theoretical \cite{Clarke87} and experimental \cite{Bentetal04} 
studies of intergranular phases in ceramic materials.  
These forces are also neglected in the present phase-field crystal study that
focuses on the structural component of the disjoining potential
due to partial crystal ordering within premelted layers.

\subsection{Sharp- and diffuse-interface theories}

In the simplest theory, a ``wet'' grain boundary is modeled 
as a thin layer of liquid sandwiched between two solid-liquid
boundaries, assumed to be sharp and straight. If only short-range
forces are present, an exponential interaction between the
interfaces is expected for large film thickness \cite{Kikuchi80}.
This suggests to write~\cite{Broughton87,Rappaz03}
\begin{equation}
\label{eq:contactpotential}
V(w)=\Delta \gamma \exp\left(-\frac w\delta\right), 
\end{equation}
where the prefactor 
$\Delta \gamma\equiv  \gamma_\mathrm{\mathrm gb}^0-2\gamma_\mathrm{\mathrm sl}$   
guarantees that the total interfacial free energy
$V(w)+2\gamma_{\mathrm sl}$ in Eq.~(\ref{Gw}) 
reduces to the energy $\gamma_\mathrm{\mathrm gb}^0$ of a ``dry"
grain boundary in the limit of vanishing liquid layer width. 
Minimization of the total excess free energy in Eq. (\ref{Gw}) with
respect to $w$, with $V(w)$ defined by Eq. (\ref{eq:contactpotential}), 
predicts that, for $\Delta \gamma>0$, the liquid layer width
vanishes for temperatures less than a ``bridging temperature'' $T_b$,
defined by $\Delta G(T_b)=-\Delta \gamma/\delta$, and increases smoothly as 
\begin{equation}
w(T)=-\delta \ln\left(-\Delta G(T)\delta/\Delta \gamma\right)~{\rm for}~T_b<T<T_m, 
\label{wet}
\end{equation}
ultimately diverging as the melting temperature $T_m$ is approached 
from below. 
For $\Delta \gamma<0$, in contrast, boundaries
remain completely dry ($w(T)=0$) for all $T$ up to a maximum  
temperature $T^*$ defined by
\begin{equation}
\Delta G(T^*)=-\Delta \gamma/\delta,\label{superheat}
\end{equation}
and are in metastable equilibrium with respect to the liquid
in the superheated range $T_m<T<T^*$.

The grain-boundary energy is generally defined as the total excess
free energy of the boundary with respect to the solid, 
or $\gamma_{\mathrm gb}(T)=G_{\mathrm{exc}}(w(T),T)$ here. Therefore, 
in the wetting case ($\Delta \gamma>0$),  
this energy is constant and simply equal to  
$\gamma_{\mathrm gb}^0$ for $T<T_b$, consistent with the requirement that
$\gamma_{\mathrm gb}(0)=\gamma_{\mathrm gb}^0$, but decreases 
for $T>T_b$ until reaching $2\gamma_{\mathrm sl}$ at the melting point where  
$w$ diverges. Substituting Eq. (\ref{wet}) into Eq. (\ref{Gw}) gives
\begin{equation}
\gamma_{\mathrm gb}(T)-2\gamma_{\mathrm sl}=
-\delta \Delta G(T) \left[1+ \ln\left(-\Delta G(T)\delta/\Delta \gamma\right)\right],\label{gben}
\end{equation}
for $T_b<T<T_m$, which has limits $\Delta \gamma$ and zero at $T=T_b$
and $T=T_m$, respectively. In contrast, for non-wetting conditions
($\Delta \gamma<0$), this theory predicts that the grain-boundary energy
retains its dry value for all temperatures:  
$\gamma_{\mathrm gb}(T)=\gamma_{\mathrm gb}^0$ for $T<T^*$.

Phase-field theories of interfacial premelting \cite{Lobkowski02,Tang06}
where solid-liquid interfaces are inherently spatially diffuse have
yielded predictions that are in part consistent with the above
picture, but also point to the possibility
of more complex premelting behaviors. The most detailed studies
have been carried out in models where
the crystal orientation is represented by a scalar field coupled
to the standard scalar phase field that measures the
local crystal disorder. For wetting conditions, those models predict 
either a smooth increase of $w$ with temperature below $T_m$, 
qualitatively similar to the behavior predicted by Eq. (\ref{wet}), 
or the existence of first-order transitions between
grain-boundary states of different 
widths~\cite{Lobkowski02,Tang06}, in analogy with
the theory of critical-point wetting~\cite{Cahn77}.
These predictions, however, depend generally on the choice
of phenomenological thermodynamic functions and parameters in those models 
that cannot be derived directly from microscopic physics, and thus are 
hard to relate to real systems. The phase-field crystal model, in 
contrast, has the advantage of removing much of the arbitrariness 
inherent in conventional phase-field theories. It explicitly
describes the dislocation structure of grain boundaries and is formulated
in terms of physical quantities, such as the liquid structure factor, that 
can be either measured experimentally or computed using MD simulations. Hence, 
this model can in principle make quantitative predictions that can be
compared to both experiments or MD simulations as demonstrated recently
for isolated solid-liquid interfaces in a bcc system \cite{Wuetal06,WuKar07}.

\subsection{Disjoining potential and layer width definitions in the phase-field
crystal model}

Before summarizing our main results, some thermodynamic
considerations relevant for the present phase-field 
crystal study are worthy of brief mention. First, while  
premelting for pure materials has traditionally
been discussed in the Gibbs ensemble of Eq.~(\ref{Gw}) 
with constant $T$, pressure $p$, and particle number $N$,
the choice of the grand canonical ensemble with constant
$T$, chemical potential $\mu$, and volume $V$, is
more suited for the PFC model where the
Helmholtz free energy is a function of the density
and simulations are carried out at fixed $V$. Thus, the
disjoining potential is defined here in terms of the excess 
of the grand potential in complete analogy with Eq. (\ref{Gw}), i.e.
it represents the total interfacial contribution of this excess minus
its asymptotic value for well-separated interfaces equal
to $2\gamma_{\mathrm sl}$. For reasons detailed below, it is simpler
to study premelting as a function of $\mu$ rather
than $T$ in the PFC model. Both are intensive variables,
and the results are expected to be equivalent. In particular, the 
departure of the chemical potential from its equilibrium value
$\mu_{\mathrm{eq}}-\mu$ in the grand canonical ensemble is analogous 
to the departure of the temperature from the melting point $T-T_m$ 
in the Gibbs ensemble, with the solid (liquid) being
stable for negative (positive) values of both quantities in 
their respective ensembles. For convenience, even though we work 
in the grand canonical ensemble with $\mu$ as control parameter,
we often refer interchangeably hereafter to temperature
and chemical potential to facilitate the  
comparison of our results to previous theories and experiments.

Second, we define the liquid layer width using a Gibbs construction.
We first determine the excess mass carried by the grain boundary,
which is simply the total mass of the bicrystal system with a 
grain boundary at fixed $\mu$ minus the mass of a single crystal 
occupying the same volume at the same $\mu$. The film thickness
$w$ is then defined by equating this excess mass to the product of
$w$ and the difference between solid and liquid densities. The advantage 
of this thermodynamic approach is that it gives a precise definition 
of $w$ that remains applicable even when 
the liquid layer is not spatially uniform along the grain boundary,
as is the case here for small misorientations. This definition
of course reduces to the standard definition of the layer width 
in the limit where the liquid layer width is much larger than
the intrinsic solid-liquid interface width ($w\gg \delta$).

\subsection{Main results}
\label{subsec:mainresults}

Let us now summarize our main results as they relate to the 
theories reviewed above. The structure and properties of 
symmetric tilt boundaries were studied as a function of 
misorientation $\theta$ for two different inclinations where the
symmetry axis (from which each crystal is rotated by $\pm \theta/2$) 
is parallel ($\phi=0$) or at a 30$^\circ$ angle ($\phi=30^\circ$)
to any of the six equivalent close-packed directions of the
hexagonal crystal.

We find that high-angle boundaries behave essentially as predicted by
the sharp-interface theory. They are dry well below the melting point 
and become uniformly wetted with a liquid layer of roughly constant
width along the boundary. The latter diverges logarithmically as the melting 
point is approached from below, consistent with a disjoining potential  
that is reasonably well approximated by 
the simple exponential form of Eq. (\ref{eq:contactpotential}).
In contrast, the behavior of low-angle boundaries in the PFC 
model is not correctly predicted by the sharp-interface theory, 
both qualitatively and quantitatively. 
The main qualitative difference is that grain boundaries in the
PFC model do not remain dry with zero width as predicted 
by this theory. They exhibit local melting around dislocations,
as previously seen in Ref. \cite{Beretal08}, and the resulting liquid pools
cause $w$ to increase smoothly with temperature (i.e., $\mu_{eq}-\mu$ in
the grand canonical phase-field crystal simulations), although
$w$ remains finite at the melting point and into the superheated range
($\mu_{eq}-\mu>0$) where these boundaries are metastable.  
At a more quantitative level, dislocation-induced premelting contributes to
the reduction in the grain-boundary energy from its low-temperature
value, which can be larger than $2\gamma_{\mathrm sl}$ even for
small misorientations, to a value less than $2\gamma_{\mathrm sl}$
near the melting point. The other factor contributing to this reduction is 
the elastic softening of the material at the melting point discussed below.

Dislocation-induced premelting of low-angle boundaries 
is reflected in the $w$-dependence of the disjoining
potential $V(w)$ that exhibits a minimum 
at a finite width $w=w_m$, which
corresponds to the equilibrium layer width
at the melting point. Therefore, this potential is repulsive for $w<w_m$
and attractive for $w>w_m$. In contrast, it is predicted to be attractive
for all $w$ in the sharp-interface theory. 
The high- and low-angle regimes can be formally distinguished 
by defining a critical misorientation $\theta_c$ such that, 
for $\theta>\theta_c$, the disjoining
potential is purely repulsive for all $w$, and, for $\theta<\theta_c$,
exhibits a minimum with short-range
repulsion and long-range attraction.  Our results suggest
that the transition between these two regimes is smooth, with the
equilibrium layer width at the melting point diverging in the 
limit where $\theta$ approaches $\theta_c$ from below, although
the nature of this divergence is hard to pinpoint precisely.
It should be emphasized that the transition does not correspond
to a sharp transition in the geometry of the grain boundary.
Rather, the critical angle falls into a range where the geometry
of the grain boundary is somewhere in between the two extremes
described above. Namely, when the melting point is approached from 
below for $\theta$ slightly above or below $\theta_c$,  
the grain boundary consists of liquid pools separated by 
``bridges'', but the distance between the dislocations is comparable
to the pool diameter, so that the pools start to overlap and the material of
the bridges is no longer fully solid.

While the main results described so far are 
ostensibly independent of inclination,  we find some
additional $\phi$-dependent features 
that require refinement of the above picture.
For the $\phi=30^\circ$ inclination, which has the simplest behavior, 
the liquid pools were always found to be 
centered around isolated dislocations for low-angle
boundaries, as seen qualitatively in Ref. \cite{Beretal08}, and to
merge progressively to form a uniform film with increasing misorientation. 
In contrast, for the $\phi=0$ inclination, which was investigated here in greater
detail, discontinuous structural transitions were seen between different
grain-boundary states. In one state, each dislocation is surrounded by
its own liquid pool. In the other state, two dislocations combine 
to share a common liquid pool. The structural transition between these
two states only occurs above a small misorientation 
well below $\theta_c$. The transition first occurs in the overheated 
states above the melting point, and shifts to lower values of
$\mu_{\mathrm{eq}}-\mu$ with increasing misorientation. Furthermore, the 
transition is hysteretic, such that two grain-boundary
states with different liquid-pool structures can coexist
over a certain range of $\mu$. The jump in $w$ at the transitions 
between different states, which measures effectively
the change in liquid fraction associated with the pairing
of liquid pools, is small. Hence, it cannot be ruled out that these
transitions would be smeared by fluctuations and not directly
observable as sharp transitions in a real system.

The most relevant aspect of our results 
for experiment is the quantitative prediction of the critical
wetting angle $\theta_c$ above which the liquid
layer width diverges at the melting point. In the sharp-interface theory, 
this angle is predicted by the standard wetting condition 
$\Delta \gamma(\theta)=\gamma_{\mathrm gb}^0(\theta)-2\gamma_{\mathrm sl}=0$, where
$\gamma_{\mathrm gb}^0(\theta)$ is taken to be the completely dry
grain-boundary energy far below the melting point. As such,
this condition predicts a value of $\theta_c$ that is much 
smaller than observed in the phase-field crystal simulations. This failure is
due to the fact that the sharp-interface theory predicts that
the grain-boundary energy is constant for nonwetting conditions. 
As noted earlier, this energy is reduced by both dislocation-induced premelting
and the elastic softening of the material at high homologous temperature. 
Therefore, one would expect a better estimate
of $\theta_c$ to be obtained by comparing $2\gamma_{\mathrm sl}$ to
a value of the grain-boundary energy at the melting point, 
$\gamma_{\mathrm gb}^m(\theta)$, which is generally much lower
than $\gamma_{\mathrm gb}^0(\theta)$ as found in a MD study 
of a tilt boundary in pure Cu \cite{Foiles94}. Note that  
$\gamma_{\mathrm gb}^m(\theta)\rightarrow 2\gamma_{\mathrm sl}$ when
$\theta$ approaches $\theta_c$ from below, and
$\gamma_{\mathrm gb}^m=2\gamma_{\mathrm sl}$ for all $\theta$ larger than $\theta_c$.

Of course, the precise determination of $\gamma_{\mathrm gb}^m(\theta)$ 
generally requires a complete solution of the problem since it 
depends on the structural details of the premelted grain-boundary structure. 
A somewhat better prediction of $\theta_c$ can nonetheless be obtained
by an estimation of the grain-boundary energy at the melting point that
takes into account the bulk elastic softening of the material and melting
around dislocations. As in previous PFC studies \cite{Eldetal02,Eldetal04},
we find that for low-angle grain boundaries $\gamma_{\mathrm gb}$ is well described 
by the Read-Shockley law \cite{ReadShockley}. The physical parameters
entering this law are the shear modulus $G$ and the dislocation core
radius $r_0$. Hence we have, for small angles, 
$\gamma_{\mathrm gb}\approx \gamma_{\mathrm RS}(\theta,G,r_0)$. 
Elastic softening and dislocation premelting are reflected
in the temperature dependence of these quantities. The shear modulus,
which can be calculated analytically in the PFC model, has large
variations: denoting by $G_0$ and $G_m$ its values at zero temperature
and at the melting point, respectively, we find typically $G_0/G_m\approx 3$.
The variation of the core radius is obtained by fitting our simulation 
data with the Read-Shockley law and describes phenomenologically
the dislocation premelting. We observe an increase of the core 
radius at the melting point by about 40\% with respect to its 
zero-temperature value.

It should be noted that, whereas data for the variation of the
elastic constants are readily available, the variation of the
core radius is a result of the premelting around dislocations
and hence difficult to quantify in experiments or MD simulations.
This suggests that it is useful to consider two successive
approximations to improve the estimate of the critical wetting
angle. If only the elastic softening is included, we can exploit
the fact that in the Read-Shockley law the grain-boundary
energy is simply proportional to the shear modulus. Therefore,
we have $\gamma_{\mathrm gb}^m(\theta)\approx \gamma_{\mathrm gb}^0(\theta)G_m/G_0$, 
and thus the modified wetting condition
$
\gamma_{\mathrm gb}^0(\theta_c)G_m/G_0\approx 2\gamma_{\mathrm sl}.
$
If, in addition, the variation of the core radius is included,
the estimate for the grain-boundary energy becomes
$\gamma_{\mathrm gb}^m(\theta)\approx \gamma_{\mathrm gb}^0\gamma_{\mathrm RS}(\theta,G_m,r_0(T_m))/
\gamma_{\mathrm RS}(\theta,G_0,r_0(0))$, where $r_0(0)$ and $r_0(T_m)$
are the values of the core radius at zero temperature and the
melting point, respectively. Inserting the explicit expression of
the Read-Shockley law yields the improved wetting condition
\begin{equation}
\gamma_{\mathrm gb}^m(\theta_c)\approx
\gamma_{\mathrm gb}^0(\theta_c)\frac{G_m}{G_0}\frac{1-\ln[2\pi\theta_cr_0(T_m)/(\alpha a)]}
                                   {1-\ln[2\pi\theta_cr_0(0)/(\alpha a)]}
\approx 2\gamma_{\mathrm sl},\label{modwet2}
\end{equation}
where $a$ is the lattice constant of the hexagonal crystal, and
$\alpha=\sqrt{3}/2$ is the distance between close-packed planes 
in the hexagonal structure, expressed in units of the lattice spacing.
Concretely, for the $\phi=0$ inclination, the phase-field crystal simulations
yield $\theta_c\approx 14^\circ$. The standard wetting condition
$\gamma_{\mathrm gb}^0=2\gamma_{\mathrm sl}$ predicts a completely
erroneous value of $\theta_c$ of about $2^\circ$. The condition
including only the elastic softening predicts $\theta_c\approx 6^\circ$, 
whereas the condition of Eq.~(\ref{modwet2}) including both effects 
yields $\theta_c\approx 10^\circ$.
The remaining discrepancy with the value from simulations reflects
the fact that the Read-Shockley law is no longer valid
when liquid pools start to overlap and the structure of the grain
boundary is no longer well described by an array of isolated dislocations, 
which is precisely the range where $\theta\approx \theta_c$. While,
therefore, even the best estimate is still of limited accuracy, it sheds 
light on several physical effects determining the critical wetting angle 
that have not been previously appreciated.

Finally, the failure of the sharp-interface theory to predict the critical 
wetting angle obviously makes this theory inadequate in predicting the 
superheated range of temperature for low-angle boundaries.
For $\phi=0$, this theory predicts that only boundaries with $\theta$ less
than $\theta_c\approx 2^\circ$ are superheated, while boundaries in the 
PFC model can be superheated for $\theta$ up to $\theta_c\approx 14^\circ$, 
with this range vanishing as $\theta\rightarrow \theta_c$. If the
computed values for the grain-boundary energy at the melting point
are used instead of the low-temperature values, the sharp-interface
prediction yields the right order of magnitude for the superheated
range for angles close to $\theta_c$, but largely overestimates this
range for low-angle grain boundaries.

The remainder of the paper is organized as follows. In
Sec.~II, we briefly review the phase-field 
crystal model and describe our numerical methods. In Sec.~III, we 
outline the procedure for obtaining the liquid layer width, the
grain-boundary energy, and the disjoining potential from
our simulation data. Our results are presented in Sec.~IV, 
and discussed in Sec.~V. Finally, concluding remarks and future 
prospects are given in Sec.~VI.

\section{Phase-field crystal model}
\label{sec:pfc}

\subsection{Basic equations and properties}
\label{sec:basic equations}

We consider the simplest PFC model defined by 
the dimensionless free energy functional~\cite{Eldetal02,Eldetal04}
\begin{equation}
\label{eq:energy}
{\cal F} = \int d\vec r \left\{
\frac{\psi}{2}[
-\epsilon + (\nabla^2 + 1)^2
] \psi
+\frac 14 \psi^4  \right\}.
\end{equation}
which is a transposition to crystalline solids of the
Swift-Hohenberg model of pattern formation~\cite{SwiHoh77}.
Furthermore, we define the dimensionless chemical potential
\begin{equation}
\mu =\frac{\delta {\cal F}}{\delta \psi} =
  -\epsilon \psi +
(\nabla^2 +1)^2\psi +\psi^3. \label{eq:muE}
\end{equation}
The dimensionless functional in Eq.~(\ref{eq:energy}) can be obtained by 
a suitable rescaling of a dimensional free-energy functional which, in 
turn, can be related to classical density-functional theory. Since
these transformations have been discussed in detail 
elsewhere \cite{Eldetal02,Eldetal04,Eldetal07,Wuetal06,WuKar07}, 
they do not need to be repeated here.

The phase diagram of this model has also been discussed 
previously \cite{Eldetal04}. However, since a precise characterization 
of the bulk phases is important for the present work, we resume here 
the main steps that are necessary to obtain the properties which are 
needed in the subsequent developments. To construct the phase diagram, 
we calculate separately the free-energy density (free energy per unit 
surface in two dimensions) as a function of the mean density $\bar\psi$ 
in the solid, denoted by $f_s(\bar \psi)$, and in the liquid, 
$f_l(\bar \psi)$, using Eq.~(\ref{eq:energy}).
Since the density is uniform in the liquid, $f_l(\bar\psi)$ 
is obtained directly from Eq.~(\ref{eq:energy}),
\begin{equation}
f_l=-(\epsilon-1)\frac{\bar\psi^2}{2}+\frac{\bar\psi^4}{4} \,.
\label{fldef}
\end{equation}
It is possible to obtain an analytical expression for $f_s$
in the one-mode approximation, in which only the contribution
of the principal reciprocal-lattice vectors is taken into
account. Then, the density for the two-dimensional hexagonal 
structure can be written~\cite{Eldetal04} as
\begin{equation}
\psi_s(x,y)=\bar\psi + A_t \left[ \cos(qx)\cos\left(\frac{q y}{\sqrt{3}}\right)
 - \frac 12 \cos\left(\frac{2q y}{\sqrt{3}}\right)\right]. 
\label{Pfc_Boston_Eq_SolHex}
\end{equation}
This solution ansatz is inserted into the free energy, 
Eq.~(\ref{eq:energy}). Integrating over a unit cell and 
minimizing the free energy with respect to $A_t$ and $q$ leads to
\begin{equation}
A_t = \frac 45 \left(\bar\psi \pm 
\frac 13 \sqrt{15 \epsilon - 36 \bar\psi^2} \right)  
\label{Pfc_Boston_Eq_At} \,,
\end{equation}
where the $\pm$ signs are for positive and negative $\bar\psi$, 
respectively, and $q=\sqrt{3}/2$. Reinserting this result into 
the free energy yields $f_s(\bar\psi)$.

The equilibrium densities of the two phases as a function
of $\epsilon$ can then be found by the common tangent construction,
which is equivalent to the requirement that the chemical
potential $\mu$ and the grand potential density 
$\omega=f-\mu\bar\psi$ must be equal in both phases,
\begin{equation}
\left. \frac{\partial f_s } {\partial \bar\psi} \right|_{\bar\psi_s} = 
\mu_s = \left.\frac{\partial f_l}{\partial \bar\psi} \right|_{\bar\psi_l} = 
\mu_l \equiv \mu_{\mathrm{eq}} \label{Pfc_Boston_Eq_CommonTangent1}
\end{equation}
\begin{equation}
\omega_s = f_s(\bar\psi_s) - \mu_s \bar\psi_s = 
\omega_l = f_l(\bar\psi_l) - \mu_l \bar\psi_l 
\,.\label{Pfc_Boston_Eq_CommonTangent2}
\end{equation}
The solution of these equations yields the equilibrium densities
as a function of temperature, $\bar\psi^{\mathrm{eq}}_s(\epsilon)$ and 
$\bar\psi^{{\mathrm{eq}}}_l(\epsilon)$.
The phase diagram of the PFC model exhibits a critical point. 
The parameter $\epsilon$ plays the role of an undercooling, that
is, higher $\epsilon$ correspond to lower temperatures.
Furthermore, the phase diagram is symmetric in $\bar\psi$
and hence exhibits two coexistence zones. We choose for all
our simulations negative values of $\bar\psi$; for this
branch of solutions the solid has a higher density than the 
liquid. Finally, for values of $\bar\psi$ close to zero, an 
additional striped (nematic) phase can exist, which is not of
importance for the present work.

The one-mode ansatz gives a good approximation for the
phase diagram as long as $\epsilon$ remains small. However, it
turns out that this approximation is not sufficient for our
purpose since we want to determine {\em excess} free energies
due to surfaces, which requires an excellent precision of the
bulk values. Therefore, we obtained the function $f_s(\bar\psi)$
from the numerical solution of the free energy minimization for 
a periodic hexagonal pattern, and used this function to perform the
common tangent construction, which leads to very precise values of
$\mu_{{\mathrm{eq}}}$, $\bar\psi^{{\mathrm{eq}}}_s$, and $\bar\psi^{{\mathrm{eq}}}_l$. 

The solid-liquid interfaces have been studied in detail
in the PFC model and in a Ginzburg-Landau model \cite{Wuetal06}. 
The main result that is important for the present work is that 
for $\epsilon$ small enough, the interfaces are smooth. That is, 
the amplitude of the density waves varies from the solid to the 
liquid  over a distance $\delta_{\mathrm sl}$ that is much larger than the spacing 
between density peaks in the solid. This makes it possible to use
a multi-scale expansion and to obtain a good approximation
for the surface tension and the order-parameter profile.
However, as for the bulk densities, this approximation is 
not precise enough for the purpose of the present work. 
Therefore, the surface tension is extracted from the 
numerical calculations as detailed below. The interface 
thickness $\delta_{\mathrm sl}$ is obtained from a fit of the density
profile $\bar\psi_x(y)$ (the density averaged over the $x$ direction, 
which is parallel to the interface) with a hyperbolic tangent, 
\begin{equation}
\bar\psi_x(y)= \frac{\bar\psi_s^{\mathrm{eq}}+\bar\psi_l^{\mathrm{eq}}}{2}+
               \frac{\bar\psi_s^{\mathrm{eq}}-\bar\psi_l^{\mathrm{eq}}}{2}
               \tanh\left(\frac{y}{\delta_{\mathrm sl}}\right), 
\end{equation}
as shown in Fig.~\ref{eps:profiles}. 
For $\epsilon=0.1$ (which is used in all the simulations 
in this work), a value of $\delta_{\mathrm sl}\approx 12.5$  is obtained.

\subsection{Numerical methods}
\label{sec:numerics}

The standard equation of motion of the PFC model is~\cite{Eldetal02,Eldetal04}
\begin{eqnarray}
\partial_t \psi &=& \nabla^2\left(\frac{\delta \mathcal{F}}{\delta \psi}\right) \nonumber \\ 
 &=&(1-\epsilon)\nabla^2\psi + 2 \nabla^4 \psi + \nabla^6\psi + \nabla^2\psi^3 
\label{Eq:Numerics:Implementation:EOM} \,,
\end{eqnarray}
which reflects the fact that the density field is a locally
conserved quantity. This equation can be efficiently solved
by using a semi-implicit pseudospectral formulation, as
detailed in Appendix A.

However, for the purpose of finding the equilibrium states,
this is not an efficient method.
The reason is that the solid and the liquid have different 
densities, which have to be adjusted to their equilibrium
values in the course of the simulation. Since
Eq.~(\ref{Eq:Numerics:Implementation:EOM}) implies that
mass is transported by diffusion only,
the equilibration time scales as the square of the system
size. Instead, a more rapid numerical scheme can be used, in 
which $\psi$ is treated as a {\em locally} non-conserved order 
parameter, while {\em global} mass conservation is ensured by a 
Lagrange multiplier. The advantage of this ``nonlocal'' method 
is that the mass can be transported faster since it can be taken 
at some space point and placed at another, as favored by the 
free energy.

The equation of motion for the nonlocal dynamics is derived
in Appendix A and can be written as
\begin{eqnarray}
\partial_t \psi &=& - \frac{\delta \mathcal{F}}{\delta \psi} + \mu \nonumber \\
&=&\left[(\epsilon-1) - 2 \nabla^2 - \nabla^4\right]\psi - \psi^3 + \mu \label{Pfc_Boston_Eq_EOM_FD}\,,
\end{eqnarray}
where the Lagrange multiplier $\mu$ is obtained as
\begin{eqnarray}
 {\mu} &=&  \frac{1}{L_xL_y} \int \left[ (1-\epsilon)\psi(\vec x) + \psi^3(\vec x)\right] d\vec x,
\label{eq:globalmu}
\end{eqnarray}
where $L_x$ and $L_y$ are the side lengths of the rectangular
simulation box. The Lagrange multiplier is the thermodynamic 
chemical potential of the system. In the scheme outlined above, 
the total mass of the system is conserved, and the chemical 
potential evolves with pseudo-time until it reaches its 
stationary equilibrium value. 

Finally, $\mu$ can also be fixed, and the constraint of global mass 
conservation released. This corresponds to a situation described
by the grand canonical ensemble, and $\mu$ is the externally imposed
chemical potential. The equilibrium state can be reached
even faster in this way since each point of the system can directly
exchange mass with the ``mass reservoir''. The equation of motion
is identical to Eq.~(\ref{Pfc_Boston_Eq_EOM_FD}), except that now 
$\mu$ is an external parameter and independent of time. This method is 
much faster than the others and will be used for almost all of the 
simulations presented below. However, it should be emphasized that 
it is not suitable to simulate isolated solid-liquid interfaces or
two-phase states within the coexistence region, since for such states 
the density $\bar\psi$ is not a unique function of $\mu$. Such states 
have therefore to be calculated with fixed total mass.

Here we neglect the effect of thermal fluctuations that is 
traditionally incorporated in the PFC model through the addition 
of a Langevin noise term in the evolution equation for the density 
field, with the amplitude of the noise determined by a standard 
fluctuation-dissipation 
relation \cite{Chandrasekhar49,Eldetal02,Eldetal04}. This choice 
is motivated by the fact that we focus primarily on computing 
quantitatively the excess interfacial free energies of dry and wet 
equilibrium grain-boundary states that correspond to stable or 
metastable free-energy minima. This requires an accurate computation 
of the free energy of the system that is readily obtained from 
a static crystal density field using the ``bare'' free-energy 
functional defined by Eq.~(\ref{eq:energy}), but that is
considerably more difficult to obtain when noise is present.
In the latter case, the additional entropy generated by the 
fluctuations of the crystal density field needs to be computed 
explicitly to obtain a ``renormalized'' free-energy functional,
which is needed to compute in a thermodynamically self-consistent 
way the disjoining potential. While such a computation is in 
principle possible (although it would require long simulations 
for statistical averaging) it appears unnecessary for the 
computation of static equilibrium properties since the 
bare free-energy functional is derived from a mean-field classical 
density-functional theory framework that already contains the effect 
of microscopic fluctuations on the atomic scale. From this fundamental 
viewpoint, noise in the PFC model can only be meaningfully defined 
in the framework of a long-wavelength hydrodynamic theory where 
it only acts on length scales larger than the correlation length. 
PFC simulations with noise in this hydrodynamic limit and without noise
should give essentially identical results as far as static equilibrium
properties are concerned. One possible exception is the case where
different grain-boundary states (corresponding to the isolated and 
paired liquid pool structures already mentioned in 
Sec.~\ref{subsec:mainresults}) are separated by small 
free-energy barriers. While such barriers are present in the 
bare free-energy landscape studied here, they could potentially 
be reduced or eliminated in the renormalized landscape due to 
frequent thermally activated transitions between these two states.

The boundary conditions have to be treated with some care.
The solid phase in the PFC model has a periodic structure and
can support strain through a variation of the wavelength with
respect to the equilibrium value. However, this variation alters
the free-energy density of the solid phase. In order to recover
the correct equilibrium values in the thermodynamic limit of
large system size, it is important to ensure that the solid
far from the grain boundaries is free from strain. Since we use
periodic boundary conditions in both $x$ and $y$ directions,
the size of the simulation box has to be carefully adjusted
to contain exactly an integer number of unstrained unit cells;
this is detailed in Appendix B.

The initial conditions used to simulate grain boundaries 
are two solid slabs which are rotated by an angle~$\Theta = \theta/2$ 
in opposite directions. The solid is created using the density 
field in the one-mode approximation~$\psi_s(x,y)$ as given in 
Eq.~(\ref{Pfc_Boston_Eq_SolHex}). The solids are initially 
separated by macroscopically large liquid films, where 
$\psi = \bar\psi_l$. Note that due to the periodic boundary 
conditions and the symmetries, there are always two equivalent grain 
boundaries in the system. To obtain ``dry'' grain boundaries, 
$\bar\psi$ (or $\mu$ in the case of grand canonical simulations) 
is chosen to be within the solid phase. Then, in the beginning of the 
simulations, the liquid rapidly solidifies and the grain boundary builds up. 
Before extracting the grain-boundary properties, the system is evolved 
for a much longer time. The approach to equilibrium can be monitored
by determining the maximum difference between the local chemical
potential given by Eq.~(\ref{eq:muE}) and the thermodynamic chemical 
potential (the Lagrange multiplier in Eq.~(\ref{eq:globalmu})
for conserved total mass, or the externally imposed value for
grand-canonical simulations).

For $\Theta=0$, a single crystal is obtained after
the liquid has disappeared. Due to the symmetry of the hexagonal 
structure, this happens also when $\Theta=30^\circ$, but the
two configurations differ. In the former case, the close-packed
rows of density peaks are aligned with the $x$ axis and hence
parallel to the initial liquid layer, whereas in the second case, 
they are aligned with the $y$ direction and hence perpendicular
to the liquid layer. Therefore, configurations with $\Theta$
close to $0$ or $30^\circ$ correspond to symmetric tilt
grain boundaries of inclination $\phi=0^\circ$ and 
$\phi=30^\circ$, respectively. Furthermore, the misorientation
is given by $\theta=2\Theta$ for $\phi=0^\circ$, but by
$\theta=60^\circ-2\Theta$ for $\phi=30^\circ$. We recover
of course the well-known fact \cite{ReadShockley} that there
are two equivalent descriptions for each grain boundary. In
the following, we will investigate the whole range of angles
$0<\Theta<30^\circ$, which includes low-angle grain
boundaries of both inclinations.

\section{Determination of the grain-boundary properties}

\subsection{General framework}

Experiments and MD simulations are mostly carried out
at constant temperature, pressure, and total number of atoms. 
Therefore, the appropriate thermodynamic potential is the 
Gibbs free energy. In the PFC model, the starting point is a 
Helmholtz free-energy functional. Simulations carried out
at fixed total mass correspond hence to constant temperature 
(here, $\epsilon$), volume, and particle number, and lead to
a minimization of the functional $\cal F$. 
In contrast, if the constraint on the
total mass is relaxed and the chemical potential is
fixed, we have constant temperature, chemical potential 
and volume, and the relevant thermodynamic potential
which is minimized by the dynamics is the grand potential,
\begin{equation}
\Omega = {\cal F} - \mu\int_V \psi.
\end{equation}
Like the Gibbs free energy, it depends on two intensive 
variables (temperature and chemical potential). We will 
formulate all the subsequent discussion in terms of the 
grand potential, and briefly discuss below how our methods 
and results can be translated to the $(N,p,T)$ ensemble
and the Gibbs free energy.

The grand potential depends on the intensive variables 
$T$ (here, $\epsilon$) and $\mu$. We will assume in the 
following developments that $T$ is kept constant and that 
only $\mu$ is varied. The motivations for this choice 
will be discussed below.
Since we have chosen the side of the PFC phase diagram
where the solid has a higher density than the liquid,
increasing the chemical potential with respect to the
coexistence value favors the solid phase. Therefore,
increasing the chemical potential is analogous to
decreasing the temperature.

The grain boundary is described as a thin film of liquid 
sandwiched between two solids, and the total grand potential 
of this two-phase system is written as
\begin{equation}
\Omega(\mu)=L_x\left[(L_y-w)\omega_s(\mu)+w\omega_l(\mu)
    +2\gamma_{\rm sl}+V(w)\right],
\label{eq:om2def}
\end{equation}
where $L_x$ is the length of the grain boundary contained in the box
(the equivalent of the total surface of grain boundary in three dimensions),
$L_y$ is the system size in the direction normal to the grain boundary,
$w$ is the thickness of the liquid film, and $\omega_s(\mu)$ and 
$\omega_l(\mu)$ are the grand potential densities of the bulk solid 
and liquid, respectively. Equation (\ref{eq:om2def}) is the direct analog
in the grand canonical ensemble of Eq. (\ref{Gw}) in the Gibbs ensemble.

As already described in Sec.~I, the last two terms 
in the brackets on the right-hand side describe the excess grand 
potential that is due to the presence of surfaces: $\gamma_{\rm sl}$ 
is the surface free energy of an isolated solid-liquid interface, 
and $V(w)$ is the {\em disjoining potential}, which 
describes the fact that two solid-liquid interfaces
start to interact when the distance between them becomes
comparable to the range of the interatomic potentials.
Since $V(w)$ describes the interaction between interfaces,
it has to tend to zero for well-separated interfaces,
$V(w)\to 0$ when $w\to\infty$. For the form of the
disjoining potential that has been assumed in the sharp-interface picture
in Eq.~(\ref{eq:contactpotential}), a distinction can be made
between ``attractive'' grain boundaries for which 
$\gamma_\mathrm{\mathrm gb}^0-2\gamma_\mathrm{\mathrm sl}<0$ (one 
grain boundary is more favorable than two solid-liquid 
interfaces), and ``repulsive'' or ``wet'' grain boundaries 
for which the opposite is true.

This terminology can be further motivated by defining the
{\em disjoining pressure} $\Pi$, frequently used in the 
physics of wetting and thin liquid films \cite{deGennes85},
\begin{equation}
\Pi = - \frac{1}{L_x} \frac{\partial \Omega}{\partial w} = 
\omega_s - \omega_l - V'(w).
\label{eq:pdef}
\end{equation}
The disjoining pressure has two contributions. The first is
of thermodynamic origin and changes sign at the melting
point. Indeed, the grand potential density can be expanded
in $\mu$ around the melting point, which yields
\begin{equation}
\omega_s - \omega_l \approx 
- \left(\bar\psi^{\mathrm{eq}}_s-\bar\psi^{\mathrm{eq}}_l\right) (\mu-\mu_{\mathrm{eq}}),
\end{equation}
where we have used the identity $\partial\omega/\partial\mu=-\bar\psi$
and the fact that $\omega_s=\omega_l$ at coexistence.
The second contribution in the disjoining pressure arises from 
the interaction of the interfaces. For the simple exponential 
form of the disjoining potential given in 
Eq.~(\ref{eq:contactpotential}), its sign depends only on the quantity 
$\Delta\gamma = \gamma_\mathrm{\mathrm gb}^0-2\gamma_\mathrm{\mathrm sl}$.

These considerations yield an alternative and quite intuitive 
picture of the phenomena already discussed in Sec.~I. 
When both contributions of the disjoining pressure are negative
($\mu>\mu_{\mathrm{eq}}$, attractive interfaces) the film thickness
vanishes ($w=0$). When both are positive ($\mu<\mu_{\mathrm{eq}}$,
repulsive interfaces), the film thickness becomes infinite.
The more interesting scenarios arise when the 
two contributions are of opposite signs: for attractive interfaces, 
metastable solids separated by a thin liquid film can exist for 
$\mu^*<\mu<\mu_{\mathrm{eq}}$. For repulsive interfaces, finite liquid films 
exist for $\mu_b>\mu>\mu_{\mathrm{eq}}$ since the repulsion between interfaces 
competes with the thermodynamic force ``pushing'' the two solids 
together. Here, $\mu^*$ and $\mu_b$ are the equivalents of the
``breaking'' and ``bridging'' temperatures $T^*$ and $T_b$ defined
in Sec. I.

We would like to point out that the notations used in
Eq.~\ref{eq:contactpotential} can easily lead to confusion because
of the use of the ``grain-boundary energy'' $\gamma_{\mathrm gb}^0$
in the expression for $\Delta\gamma$.
Indeed, the grain-boundary energy of {\em any} grain boundary,
be it ``dry'' or ``wet'', is defined as the total excess grand
potential per unit length of grain boundary with respect to 
a {\em single-phase solid}. Therefore, the grain-boundary 
energy is
\begin{eqnarray}
\gamma_{\mathrm gb}(\mu) & = & \frac{\Omega(\mu)-L_xL_y\omega_s(\mu)}{L_x} \nonumber \\
      & = & (\omega_l-\omega_s)w_{\mathrm{eq}}(\mu) \nonumber \\
      &   & \mbox{} + 2\gamma_{\mathrm sl} + V(w_{\mathrm{eq}}(\mu)),
\label{eq:Vggbrelation}
\end{eqnarray}
where the equilibrium film thickness for given chemical potential,
$w_{\mathrm{eq}}(\mu)$, is obtained from the condition that $w_{\mathrm{eq}}$ minimizes
the grand potential (which corresponds to a vanishing disjoining
pressure),
\begin{equation}
V'(w_{\mathrm{eq}}(\mu)) = \omega_s(\mu) - \omega_l(\mu).
\label{eq:eqfilmthickness}
\end{equation}
It can be easily seen that $\gamma_{\mathrm gb} =
\gamma_{\mathrm gb}^0$ only when $w=0$.
It should be emphasized that Eq.~(\ref{eq:Vggbrelation})
is completely general, and not limited to the special case
of an exponential disjoining potential. This relation, which
shows that the grain-boundary energy and the disjoining potential 
are not independent, can actually be exploited to determine the 
disjoining potential, as will be detailed below.

\subsection{Liquid film thickness}

To proceed, we need a way to extract the liquid film
thickness from our simulation data. When the two 
solid-liquid interfaces are well separated, it is easy 
to define a film thickness by the distance between
the midpoints of the diffuse interfaces. However, this
definition becomes obsolete when the diffuse interfaces
overlap. Another definition is needed; we choose here to
use a Gibbs construction.

When the liquid film is macroscopically large (that is, the
separation between the two solid-liquid interfaces is much
larger than the intrinsic interface width), the interfaces 
do not interact (the disjoining pressure vanishes) and we 
are in the case of two-phase coexistence, which implies 
that $\mu=\mu_{{\mathrm{eq}}}(\epsilon)$.
The volume fractions of liquid and solid are related to
the total mass of the system by the lever rule. For a 
one-dimensional system of length $L_y$ and a film of 
thickness $w$, we have
\begin{equation}
\bar\psi L_y = \bar\psi_l w + \bar\psi_s (L_y-w).
\label{eq:leverrule}
\end{equation}
with $\bar\psi_l = \bar\psi_l^{\mathrm{eq}}$ and  
$\bar\psi_s = \bar\psi_s^{\mathrm{eq}}$.

\begin{figure}
\begin{center}
\includegraphics[width = 8 cm]{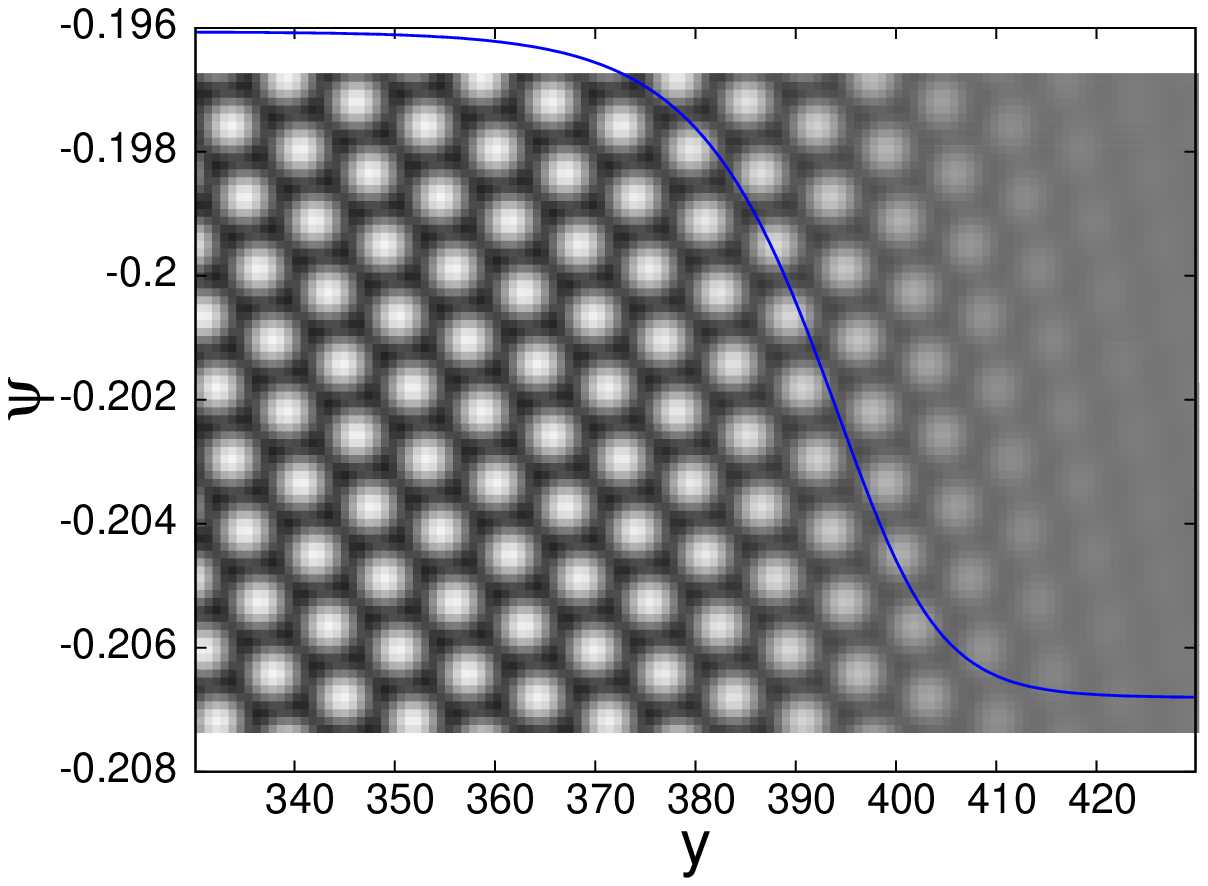}\\
\includegraphics[width = 8 cm]{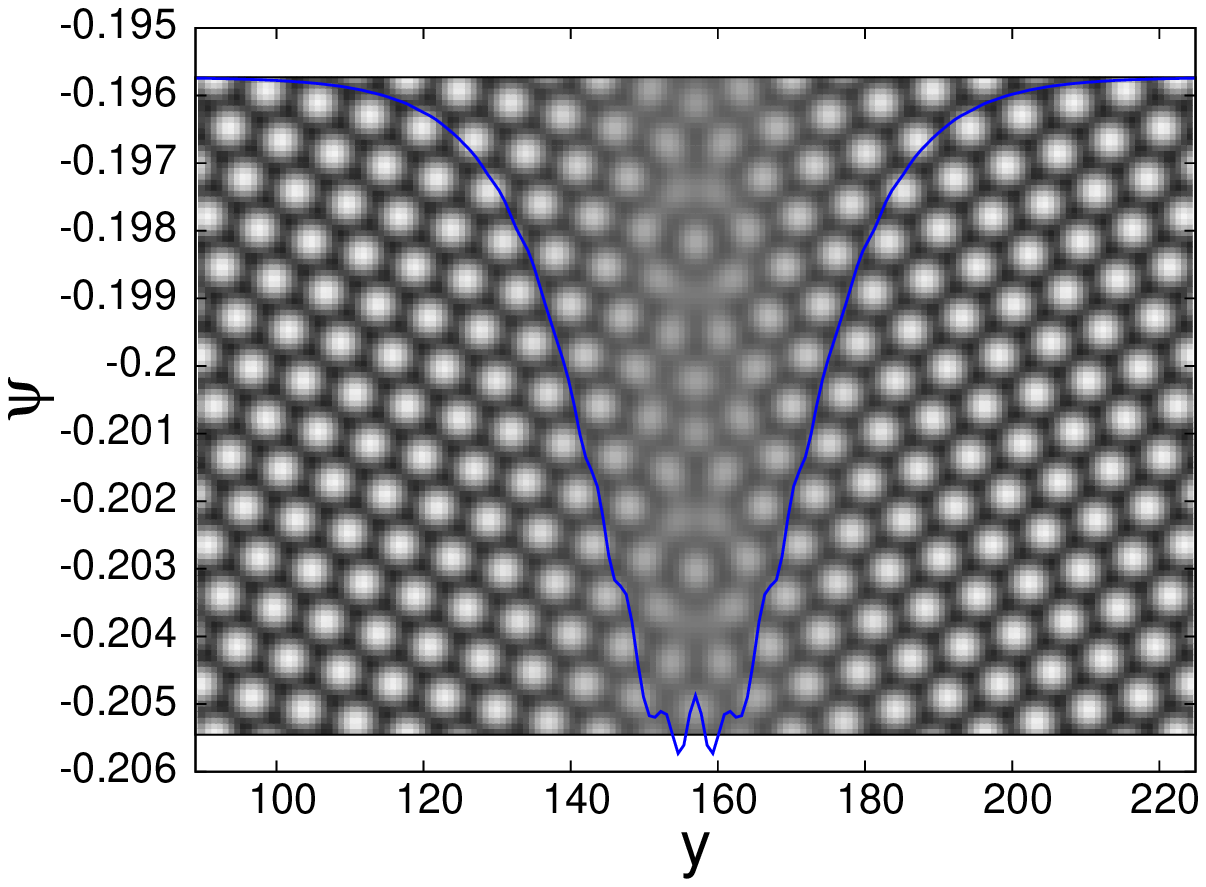}\\
\includegraphics[width = 8 cm]{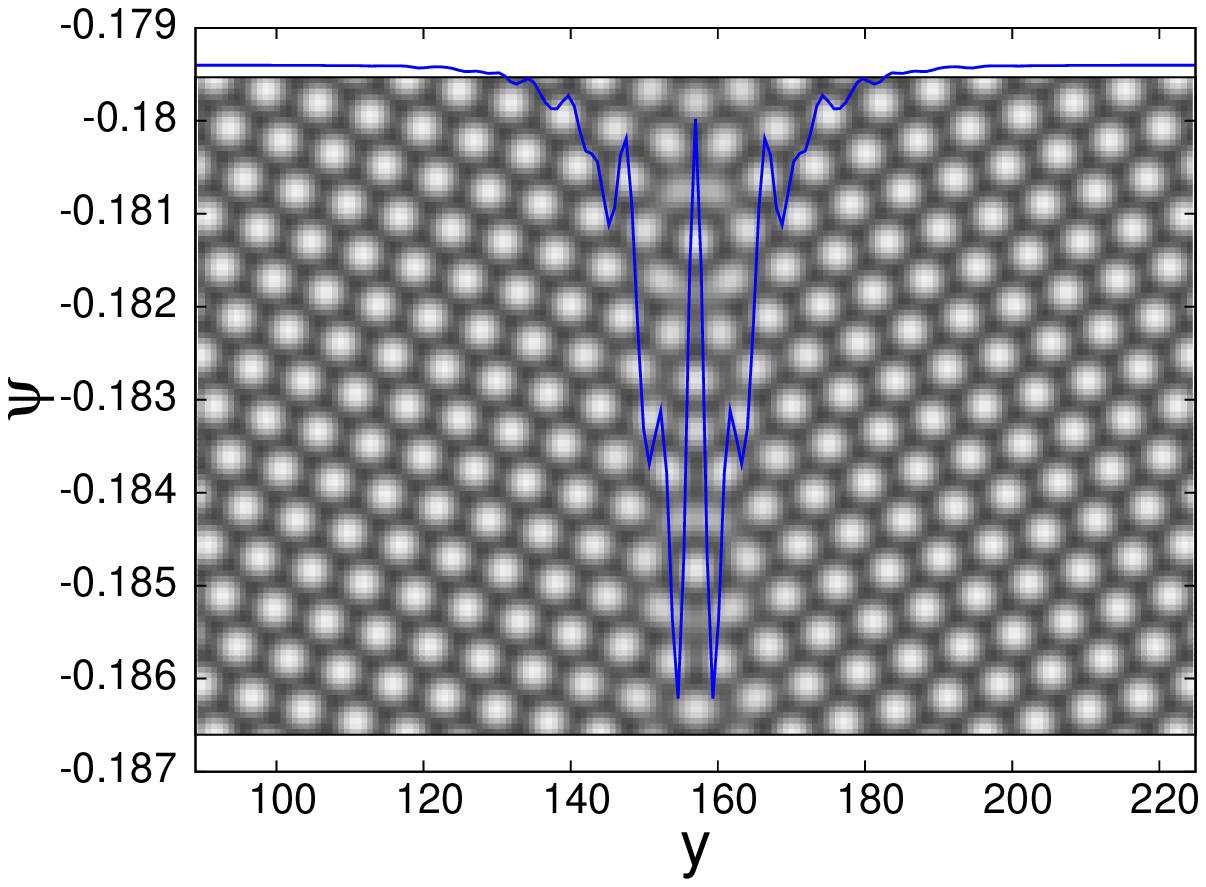}
\caption{
\label{eps:profiles}(Color online)
Density profiles of solid-liquid interface and grain boundaries. 
The complete two-dimensional density field $\psi(x,y)$ is shown.
The superimposed line gives, for any point along the direction normal
to the grain boundary, the value of the density averaged over 
the direction parallel to the grain boundary
($\bar\psi_x(y)=(1/L_x)\int \psi(x,y)\;dx$).
Top: Solid-liquid interface, $\theta = 21.8^\circ$.
Middle: Solid-solid interface close to the melting point, 
$\bar\psi=-0.1980$, $\theta=32.2^\circ$.
Bottom:  Solid-solid interface far from the melting point, 
$\bar\psi=0.180$, $\theta = 32.2^\circ$.}
\end{center}
\end{figure}

This is no longer valid when the interfaces interact:
the disjoining pressure modifies the
equilibrium chemical potential. However, volume fractions
can still be defined starting from the consideration that
the solid is a bulk phase which occupies a macroscopic volume. 
Consequently, the relation between its density and
chemical potential is the same as that for a homogeneous
solid. In contrast, the ``liquid'' film is microscopic, and
hence this region does not have the properties of a bulk
liquid. This is illustrated in Fig.~\ref{eps:profiles}, 
where we show a numerically calculated equilibrium state together 
with a plot of the density averaged over the direction parallel 
to the grain boundary. The density exhibits a ``dip'' and
approaches the value of the liquid when the film thickness
is relatively large.  It exhibits an oscillatory behavior
for more ``dry'' grain boundaries, but the average density
in the grain-boundary region is still different from the
one in the bulk solid.

This density change in the grain-boundary region can be
exploited to define a film thickness. An {\em excess mass} 
per unit length of grain boundary can be defined by subtracting 
the mass of the homogeneous solid at the same chemical potential 
from the actual mass contained in the system,
\begin{equation}
\psi_{{\mathrm{exc}}}(\mu) = L_y\left[\bar\psi - \bar\psi_s(\mu)\right].
\label{eq:excessmass}
\end{equation}
Furthermore, it is easy to obtain the density of a bulk
liquid at the same chemical potential, $\bar\psi_l(\mu)$
from the curve of $f_l(\bar\psi)$. Then, the film thickness
can be defined by the requirement that the density difference
of the bulk phases times the film thickness is equal to the
excess mass,
\begin{equation}
w\left[\bar\psi_l(\mu)-\bar\psi_s(\mu)\right] = \psi_{{\mathrm{exc}}}(\mu).
\label{eq:filmthickness}
\end{equation}
Putting these two equations together, we obtain again the
lever rule, but this time with $\bar\psi_s(\mu)$ and
$\bar\psi_l(\mu)$ instead of the coexistence values.
With this definition, the film thickness can be extracted
with good precision from simulations either at fixed total
mass ($\mu$ is measured in the simulation) or at fixed
chemical potential (the total density $\bar\psi$ is measured).

\subsection{\label{SubSec:Numerics:ContactPontential}Grain-boundary energy
and disjoining potential}

It turns out that the direct numerical determination of the 
grain-boundary energy requires some care. It is defined as the excess 
of grand potential. Contrary to the mass excess defined above, 
the straightforward method of subtracting the grand potential 
of a homogeneous bulk solid from the total grand potential of 
the simulated system leads to large numerical errors. This is
most likely due to the evaluation of the gradient contributions
in the free energy. A more precise method is to exploit the 
dependence on system size. By dividing Eq.~(\ref{eq:om2def}) 
through $L_xL_y$ and using the definition of the grain 
boundary energy, we obtain that the total grand potential 
density varies with system size at fixed chemical potential as 
$\omega=\omega_s+\gamma_{\mathrm gb}/L_y$.
The grain-boundary energy can therefore be obtained from a 
plot of $\omega$ versus the inverse system size.

\begin{figure}
\begin{center}
\includegraphics[width = 8 cm]{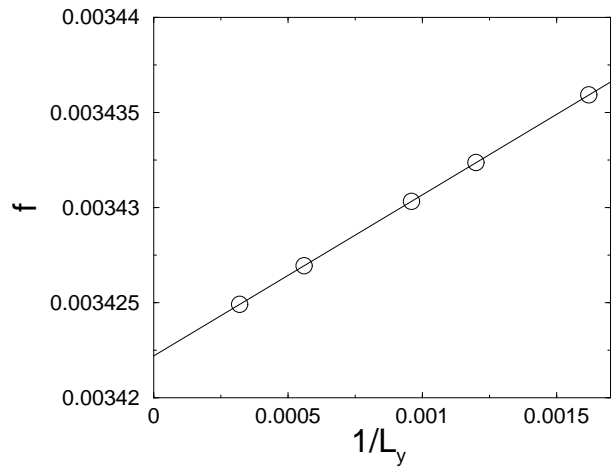}
\caption{Symbols: Free energy densities $f$ of systems with the 
same misorientation and average densities $\bar\psi$ but different 
lengths $L_y$ perpendicular to the grain boundary, plotted versus
$1/L_y$. Line: Linear fit to the data. The slope gives twice the 
grain-boundary energy, where the factor of 2 is due to the fact 
that in periodic systems there are always two grain boundaries. 
In this example, $\epsilon=0.1$, $\bar\psi=-0.1$,and the 
misorientation is $\theta=6^\circ$.
\label{eps:ffit}}
\end{center}
\end{figure}

A second, and slightly simpler way, to obtain the grain
boundary energy is to perform simulations at a fixed total
density $\bar\psi=\bar\psi_0$ and to use the free energy 
density, which can be directly obtained from the simulations.
Indeed, since the density in the grain boundary is 
different from that in the bulk, for a fixed total density 
and length of grain boundary, the bulk density in the 
solid $\bar\psi_s$ (and therefore also the chemical 
potential) vary with the system size. In the limit 
$L_y\to \infty$, the bulk density $\bar\psi_s$ tends
to $\bar\psi_0$ and the chemical potential tends to the value
corresponding to a solid at that density. Using the
result obtained above, $\omega=\omega_s+\gamma_{\mathrm gb}/L_y$,
and the definition $\omega=f-\mu\bar\psi$ we obtain
\begin{equation}
f = f_s(\bar\psi_s) - \mu (\bar\psi_s - \bar\psi_0) 
  + \frac {\gamma_{\mathrm gb}}{L_y}.
\end{equation}
Expanding $f_s$ in $\bar\psi$ around $\bar\psi_0$, using that
$\partial f/\partial \bar\psi = \mu$, and inserting the result
in the above equation, all the terms involving $\mu$ cancel out, 
and finally we obtain
\begin{equation}
f = f_s(\bar\psi_0) + \frac {\gamma_{\mathrm gb}}{L_y}.
\label{eq:fvsLy}
\end{equation}
Therefore, we determine the excess grand potential by
performing simulations at fixed $\bar\psi_0$ and calculating
the free energy directly from the free energy functional.
Note that $\gamma_{\mathrm gb}$ depends on $\mu$ and hence also
varies with system size. However, this gives rise to terms
in $f$ that are of order $1/L_y^2$ and should therefore
be small. A plot of the total free-energy density versus 
$1/L_y$, as shown in Fig.~\ref{eps:ffit},
is indeed well fitted by a straight line. Therefore, we can
extract the slope and intercept, which correspond to
$\gamma_{\mathrm gb}$ and to the free-energy density in 
the thermodynamic limit, respectively. A numerical error 
can also be estimated from the fit if more than two 
different lengths are simulated. The same procedure is 
also used to determine the solid-liquid surface tensions.
For this, it is sufficient to choose an average density
which leads to macroscopically large liquid films.

The disjoining potential can then be obtained in two ways. We 
remark that $\omega_l - \omega_s$ is a function of $\mu$ 
only, and thus $V'(\mu)$ is a known function of $\mu$ which
depends only on bulk thermodynamics. The extraction of the liquid 
layer thickness from the simulations yields $w_{\mathrm eq}(\mu)$.
The two can be combined to yield $V'(w)$ which can then
be integrated to $V(w)$. Alternatively, once the grain-boundary 
energy is calculated, Eq.~(\ref{eq:Vggbrelation}) can 
be used to obtain $V(\mu)$, which can again be combined with 
$w(\mu$) to yield $V(w)$. While this second approach avoids
a numerical integration, it is also more costly since for
each value of the grain-boundary energy several simulations
with different system sizes have to be performed.

\subsection{\label{subsec:consistency}Thermodynamic consistency}

It is useful to comment here on two important points with 
regards to the thermodynamics of interfaces and grain boundaries.
The first is that with our definition of the film thickness, the
disjoining potential and the
grain-boundary energy are entirely thermodynamically consistent.
To show this, let us first remark that Eq.~(\ref{eq:Vggbrelation})
for the grain-boundary energy formally depends on two
variables, $\mu$ and $w$. Taking the differential of this
equation, we find
\begin{eqnarray}
d\gamma_{\mathrm gb} & = & \left[\omega_l-\omega_s + V'(w)\right]dw +
            w\left[\frac{\partial\omega_l}{\partial\mu}-
                   \frac{\partial\omega_l}{\partial\mu}\right]d\mu \nonumber\\
  & = & -\Pi dw - \psi_{\mathrm{exc}} d\mu,
\end{eqnarray}
where we have used the definitions of $\Pi$ and $w$ and the
fact that $\partial\omega/\partial\mu=-\bar\psi$ to obtain
the second equality. It is clear that the film
thickness $w$ plays, for the excess free energy, the same
role as the volume in bulk thermodynamics. Furthermore,
since at equilibrium the disjoining pressure vanishes,
$\Pi=0$, the variation of the grain-boundary energy is
consistent with the fundamental definition of interfacial
excess quantities \cite{CahnReview}. Indeed, since $\gamma_{\mathrm gb}$ 
is an excess of grand potential, we can write
\begin{equation}
\gamma_{\mathrm gb} = f_{\mathrm{exc}} - \mu \psi_{\mathrm{exc}},
\end{equation}
where $f_{\mathrm{exc}}$ is the excess free energy;
differentiation with respect to $\mu$ yields
\begin{equation}
\frac{\partial\gamma_{\mathrm gb}}{\partial\mu} = -\psi_{\mathrm{exc}}.
\label{eq:thermintegration}
\end{equation}
As a corollary, once a value of $\gamma_{\mathrm gb}$ is known for
a single value of $\mu$, the curve $\gamma_{\mathrm gb}(\mu)$ can be 
obtained by integrating the function $-\psi_{\mathrm{exc}}(\mu)$ 
extracted from the simulations. Furthermore, formally the 
grain-boundary energy can also be obtained by keeping $\mu$ fixed
and integrating the ``mechanical work'' $-\Pi dw$ over $w$,
noticing that the disjoining pressure is non-zero if $w\neq w_{\mathrm{eq}}$.
This procedure, however, cannot be carried out in practice
since the configurations with $\Pi\neq 0$ are not equilibrium
states and hence cannot be obtained in simulations.

The second remark concerns the generalization of our definition
and procedures to other variables and ensembles. The various 
relationships between $w$, $V(w)$, and $\gamma_{\mathrm gb}$ obtained
above all make use of the fact 
that the film thickness has been defined by a Gibbs construction
using the interface excess of the density, which is the extensive 
quantity conjugate to the externally controlled intensive
variable $\mu$. Equivalent constructions can of course be
performed with other pairs of variables. For instance, in their
lattice-gas study, Kikuchi and Cahn kept the chemical potential
constant and varied the temperature, while they defined the 
thickness of the liquid layer by the excess of entropy \cite{Kikuchi80}.
Similarly, in the $(N,p,T)$ ensemble, a film thickness can
be defined {\em via} the excess entropy for varying temperature,
or {\em via} the excess volume for varying pressure. Since,
in this ensemble, the volume is no longer constant, instead of 
volume densities as above, quantities normalized by the particle 
number have to be used. Nevertheless, following the ideas in 
Ref.~\cite{CahnReview} to treat this change in normalization, all 
the relations given above can be translated without difficulties.

A more complex situation arises if both the chemical potential 
and the temperature (here, $\epsilon$) are allowed to vary.
For clarity of exposition, we use in the remainder of this
subsection the temperature $T$ instead of the dimensionless 
quantity $\epsilon$.
The definition of the grain-boundary energy and its variation become
\begin{eqnarray}
\gamma_{\mathrm gb} & = & e_{\mathrm{exc}} - Ts_{\mathrm{exc}} - \mu\psi_{\mathrm{exc}} \\
d\gamma_{\mathrm gb} & = & -\Pi dw - s_{\mathrm{exc}} dT - \psi_{\mathrm{exc}} d\mu,
\end{eqnarray}
respectively, where $e_{\mathrm{exc}}$ and $s_{\mathrm{exc}}$ are the interfacial
excesses of the internal energy and the entropy, respectively. 
Since $\gamma_{\mathrm gb}$ as well as all the other excess quantities
are defined as excesses with respect to a bulk thermodynamic 
potential that depends on $\mu$ and $T$, they are all state
functions, that is, unique functions of the
two intensive variables $\mu$ and $T$. The same is
true of the film thickness $w$, which is defined through
an interfacial excess quantity. 

In contrast, the disjoining
potential is {\em not} a state function. This is easy to
see when considering the equilibrium condition for the
film thickness, Eq.~(\ref{eq:eqfilmthickness}): its right-hand
side, $\omega_s - \omega_l$, now depends on the two
variables $\mu$ and $T$, which implies that $V'(w)$
has the same dependency. If this is to be integrated to a
function of a single variable $w$, a direction in the space
spanned by $\mu$ and $T$ has to be specified. Another
way to state the same fact is to remark that in 
Eq.~(\ref{eq:Vggbrelation}), $\gamma_{\mathrm gb}$ depends on the
two independent variables $\mu$ and $T$, whereas
the ``reference value'' $2\gamma_{\mathrm sl}$ depends only on
one independent variable, since $\gamma_{\mathrm sl}$ is only
defined on the coexistence line in the phase diagram, 
$\mu_{\mathrm{eq}}(T)$. For a given point away from this line, 
where $\gamma_{\mathrm gb}$ is still defined, $V$ can be
defined only if a reference point on the coexistence line
is specified. This amounts to specifying the path in the
state space that is to be followed. In the developments
above, we have supposed a particularly simple path, namely,
a constant value for one of the variables. It would be
possible to extract from our PFC model disjoining potentials 
at constant pressure or at constant density: for both cases,
the bulk equation of state for the solid (which can be obtained
from $f_s(\mu,T)$) fixes a relation between $\mu$ and $T$, 
and $V'(w)$ can be integrated along this path. Note, however,
that this procedure requires the calculation of both the
excess mass and the excess entropy.

In summary, the definition of the disjoining potential is only 
meaningful if the corresponding path in thermodynamic state space 
is specified, and the knowledge of a single disjoining potential
yields only a partial knowledge about the premelting transition. 
The more general quantity is the grain-boundary energy, which is 
the thermodynamic potential for the interfacial excess quantities. 
If its dependence with respect to the two intensive variables 
is known, all the possible disjoining potentials can be easily 
extracted using Eq.~(\ref{eq:Vggbrelation}).

\subsection{Choice of simulation parameters}

Having the discussion of Sec.~\ref{subsec:consistency} in mind,
we need to choose a particular path in the state space to
investigate the disjoining potential. It would be possible
to approach the melting transition from the solid side for
a fixed chemical potential by decreasing $\epsilon$. However,
extracting the excess entropy is far more delicate than
extracting the excess mass. Therefore, in the following we
prefer to keep $\epsilon$ fixed to $0.1$ (a value that has
been obtained for the equilibrium solid-liquid interfaces
in pure iron with body-centered-cubic crystal ordering 
\cite{Wuetal06,WuKar07}) and to explore the melting 
transition by varying the chemical potential $\mu$.

For the subsequent presentation of the results, we will use 
the following rescaled variables:
\begin{equation}
\Delta = \frac{\bar\psi-\bar\psi_s^{{\mathrm{eq}}}}
                {\bar\psi_s^{{\mathrm{eq}}}-\bar\psi_l^{{\mathrm{eq}}}}.
\end{equation}
This corresponds to a supersaturation. Furthermore, we define
a scaled chemical potential by
\begin{equation}
u  = \frac{\mu_{{\mathrm{eq}}}-\mu}{\left.\frac{\partial\mu}
       {\partial\bar\psi_s}\right|_{\bar\psi_s^{\mathrm{eq}}}
       (\bar\psi_s^{{\mathrm{eq}}}-\bar\psi_l^{{\mathrm{eq}}})},
\label{eq:udef}
\end{equation}
where the sign is chosen to stress the analogy between $u$
and a temperature \cite{remark2}: for $u<0$ ($u>0$), the solid 
(liquid) is the favored phase. For $\mu$ close to the coexistence 
value, the numerator can be expanded in $\bar\psi$, which 
yields $u\approx-\Delta$. We list in Table~\ref{table1}
the values of all the quantities needed for this scaling.
Furthermore, we will often rescale the grain-boundary energy
by $2\gamma_{\rm sl}$, and lengths by $a$, the lattice spacing
of the hexagonal crystal. The values of these quantities are
also given in Table~\ref{table1}.

\begin{table}
\caption{Numerical values of various quantities needed to
scale the density, chemical potential, grain-boundary energy,
and lengths, for $\epsilon=0.1$.
\label{table1}}
\begin{tabular}{lcc}
Quantity & Symbol & Value \\
\hline
Solid density at coexistence & $\bar\psi_s^{\mathrm{eq}}$ & $-0.19696406$ \\
Liquid density at coexistence & $\bar\psi_l^{\mathrm{eq}}$ & $-0.2068060$ \\
Chemical potential at coexistence & $\mu_{\mathrm{eq}}$ & $-0.19497015$ \\
Slope of the curve $\mu$ versus $\bar\psi_s$ & $\partial\mu/\partial\bar\psi_s$ & $0.731218$ \\
Solid-liquid surface tension ($\times 2$) & $2\gamma_{\rm sl}$ & $0.00192$ \\
Lattice constant & $a$ & $\frac{4\pi}{\sqrt{3}} \approx 7.2552$ \\
\end{tabular}
\end{table}

\section{Results}

\subsection{\label{SubSection:ResultsStructure}Structure of the grain boundaries}

\begin{figure}
\begin{center}
\includegraphics[width = 3 cm]{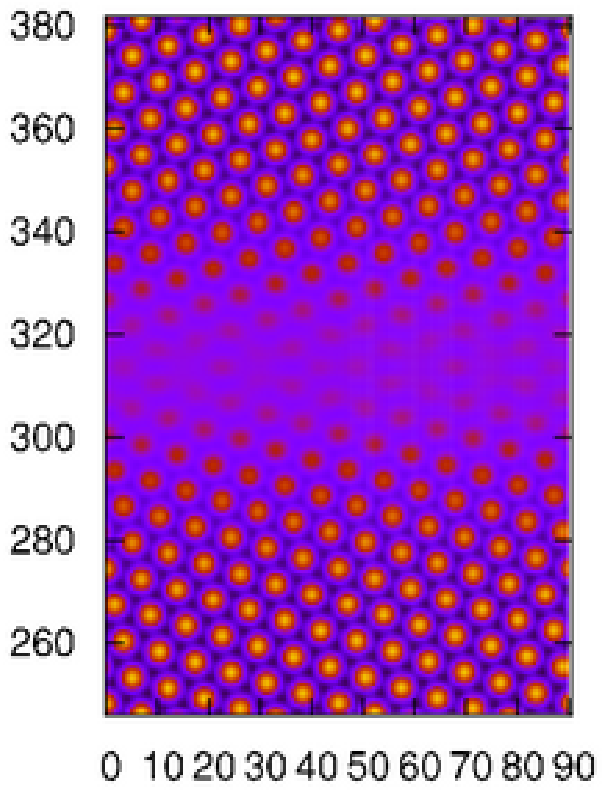}
\hspace{5mm}
\includegraphics[width = 3 cm]{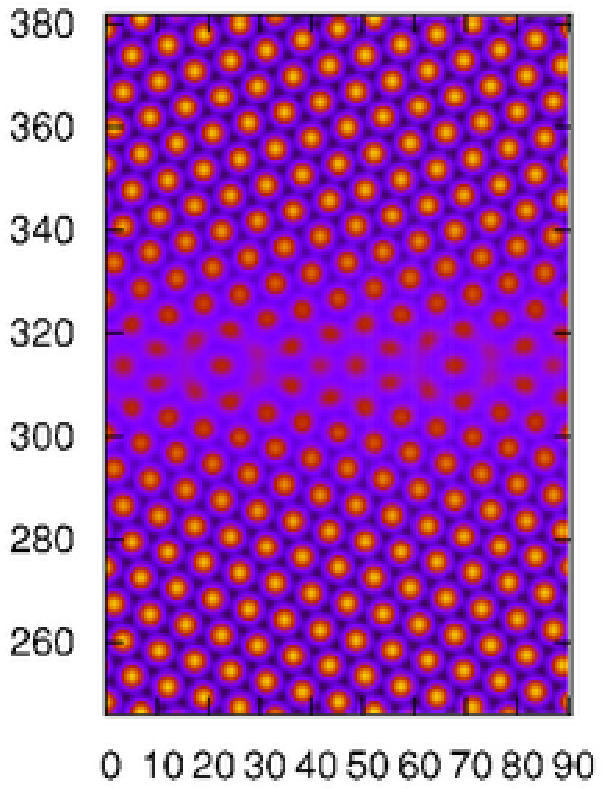}\\
\vspace{2mm}
\includegraphics[width = 3 cm]{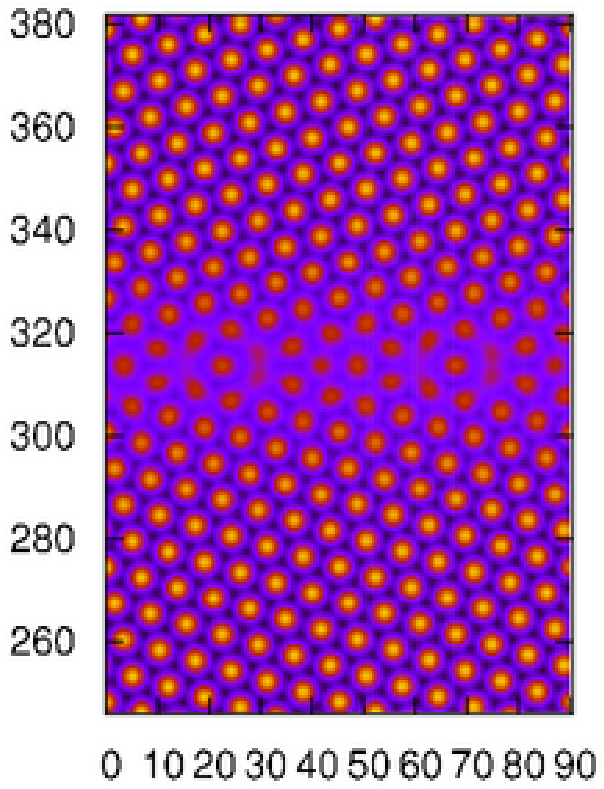}
\hspace{5mm}
\includegraphics[width = 3 cm]{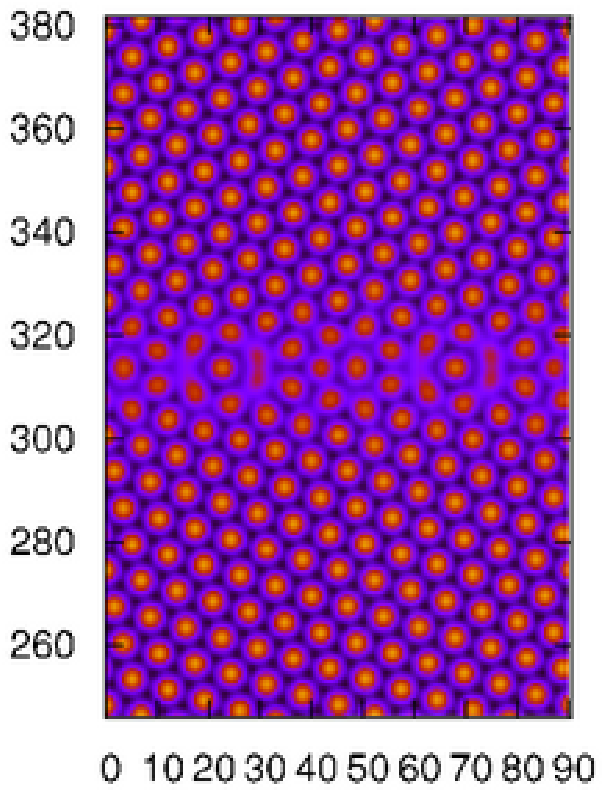}\\
\vspace{2mm}
\includegraphics[width = 3 cm]{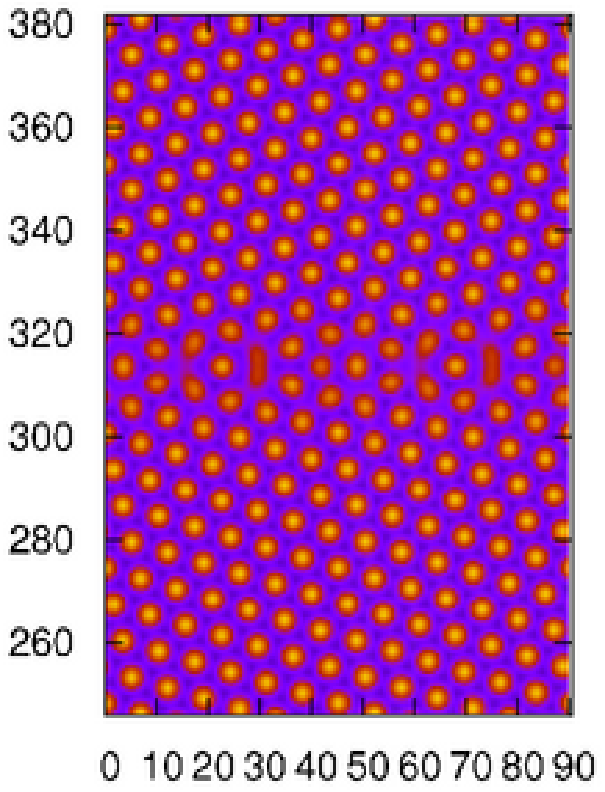}
\hspace{5mm}
\includegraphics[width = 3 cm]{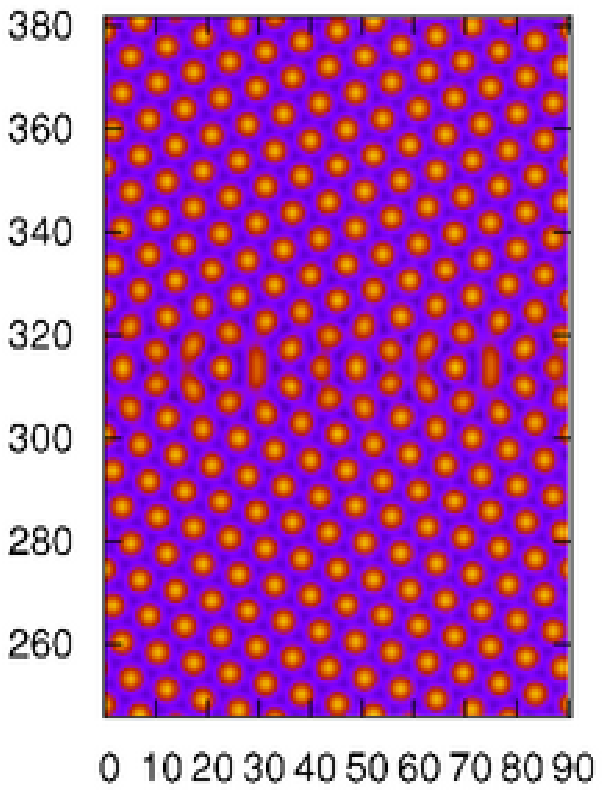}
\caption{(Color online) Snapshots of a high-angle grain boundary 
with $\theta = 32.2^\circ$ for different values of $u$, which 
increase from bottom right to top left
(see text and Fig.~\protect\ref{eps:Xexample} for details). 
The ``liquid'' forms a rather homogeneous film. Only part
of the simulation box is shown.
\label{eps:HighAngle}}
\end{center}
\end{figure}

\begin{figure}
\begin{center}
\includegraphics[width = 3.5 cm]{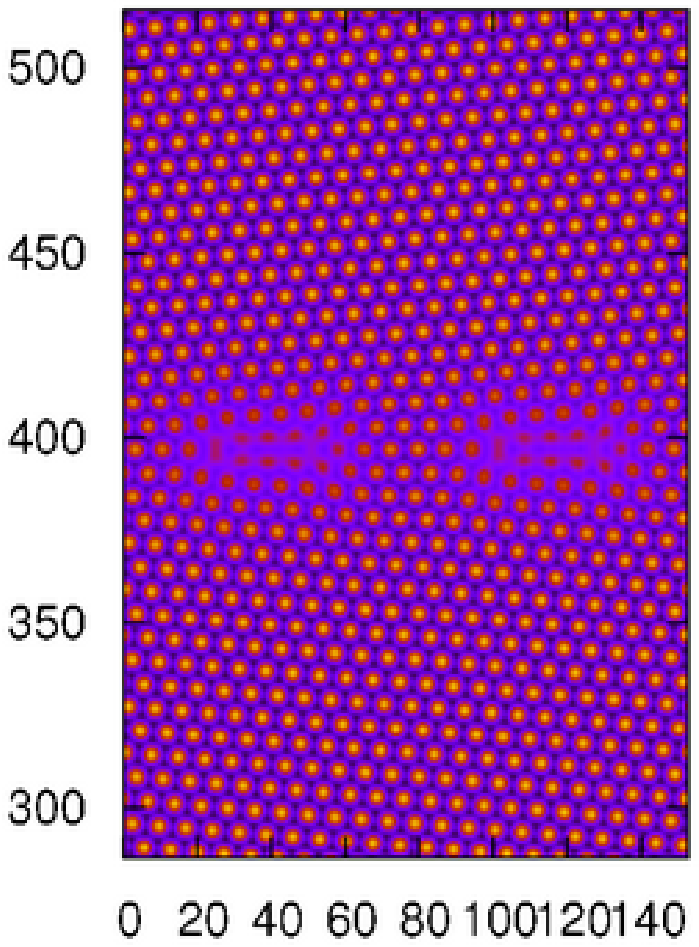}
\hspace{5mm}
\includegraphics[width = 3.5 cm]{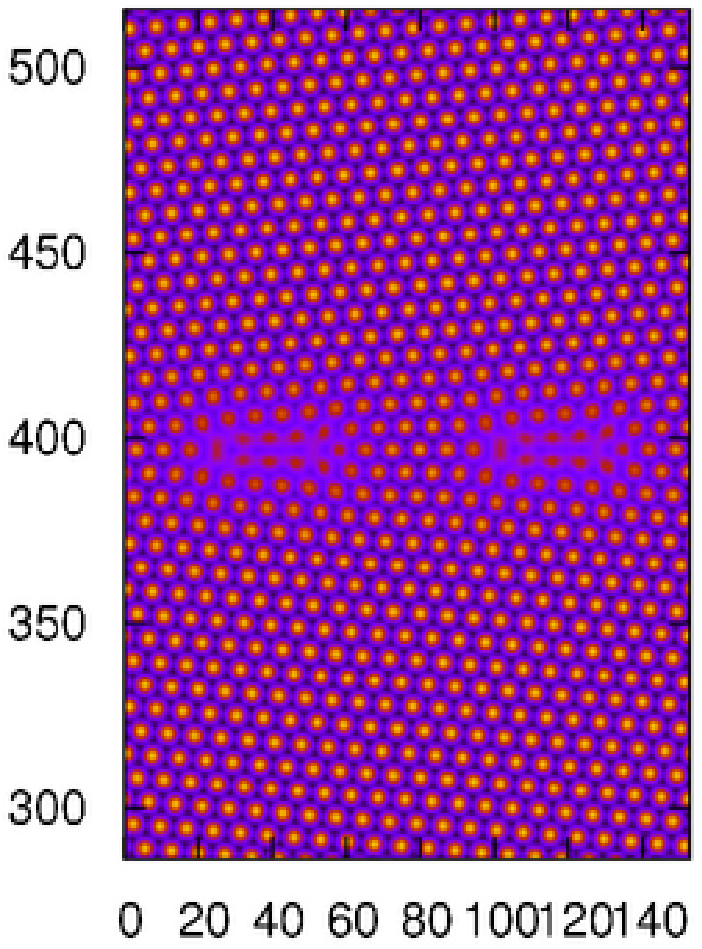}\\
\vspace{2mm}
\includegraphics[width = 3.5 cm]{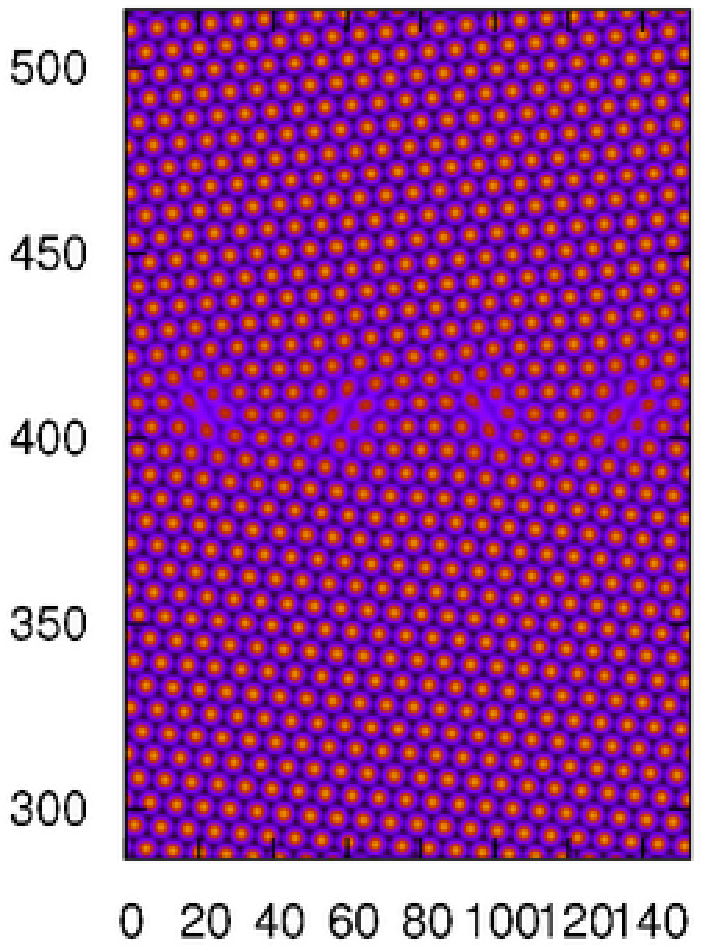}
\hspace{5mm}
\includegraphics[width = 3.5 cm]{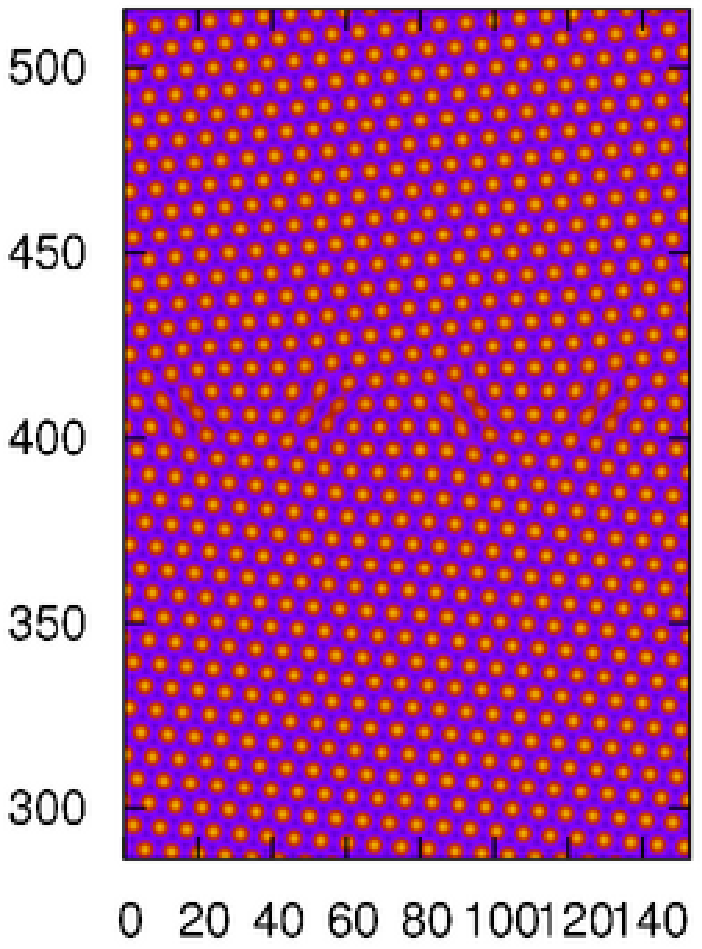}
\caption{(Color online) Snapshots of a low-angle grain boundary 
with $\theta = 9.4 ^\circ$ 
for different values of $u$, which increases from bottom right
to top left (see text and Fig.~\protect\ref{eps:Xexample} for details).
The grain boundary consists of individual dislocations and
undergoes a structural transition. Only part of the simulation
box is shown.
\label{eps:LowAngle}}
\end{center}
\end{figure}

\begin{figure}
\begin{center}
\includegraphics[width = 8 cm]{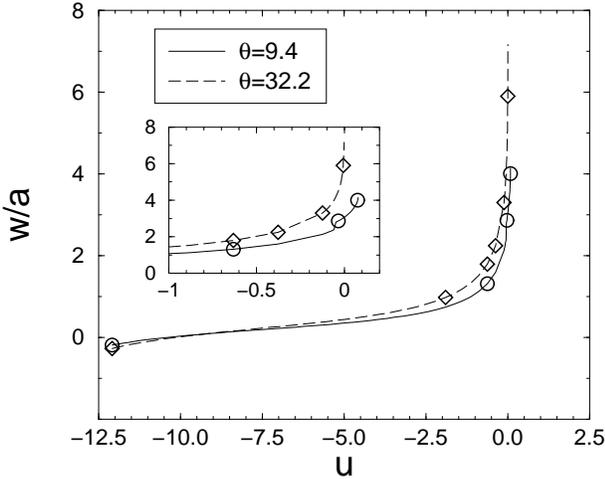}
\caption{
Ratio of film thickness $w$ to lattice spacing $a$ as a function of scaled chemical 
potential $u$ for two different grain boundaries. The inset shows a blowup of the 
vicinity of the melting point. The symbols mark the states that are depicted in 
the snapshot pictures in Figs.~\protect\ref{eps:HighAngle}
and \protect\ref{eps:LowAngle}.
\label{eps:Xexample}}
\end{center}
\end{figure}

Our simulations reveal that there is a strong difference
in behavior between high-angle and low-angle grain boundaries.
In order to illustrate first a few important features, we show in 
Figs.~\ref{eps:HighAngle} and~\ref{eps:LowAngle}
snapshot pictures of a high-angle and a low-angle grain boundary
of inclination $\phi=0^\circ$, for different values of $u$. Furthermore, 
we plot in Fig.~\ref{eps:Xexample} the curves of film thickness $w$
versus scaled chemical potential $u$ corresponding to the same 
two grain boundaries. We have checked that canonical and 
grand-canonical simulations (fixed total mass and fixed 
chemical potential, respectively) yield identical results 
for the film thickness and the grain-boundary structure.

For both high-angle and low-angle grain boundaries, the film thickness
becomes negative far below the melting point. Indeed, formally,
since the film thickness is defined {\em via} an excess mass,
it does not need to remain positive. A negative film thickness
corresponds to an accumulation of mass in the grain boundary
instead of the depletion observed in Fig.~\ref{eps:profiles}. 
When $u$ is increased, the film thickness becomes positive,
but remains small until the vicinity of the melting point is
reached. For the high-angle grain boundary, the film thickness 
then increases rapidly and diverges as the melting point is
approached from below; this is the behavior expected for a
repulsive grain boundary. In the snapshot pictures, it can be 
seen that the liquid film is rather homogeneous, that is, it 
has approximately the same width at every point. 

The low-angle grain boundary depicted in Fig.~\ref{eps:LowAngle}
consists of individual dislocations separated by distances 
that are larger than a few lattice spacings.
Here, the ``liquid'' first appears in the form of ``pools''
around the dislocations, and there is no homogeneous film 
of liquid. Furthermore, as the melting point is approached,
a structural transition occurs: the dislocations form pairs;
that is, two dislocations join and are surrounded by a common
liquid pool. This transition is accompanied by a jump in the 
film thickness $w$. Furthermore, this structure can be ``overheated'';
that is, such states exist even for $u>0$, which indicates
an attractive grain boundary. The pools grow in size, 
thus reducing the strength of the ``bridges'' of solid.
At a critical overheating, the bridges break, and the whole 
system becomes liquid.

Most of the features described above -- dependence of the film
thickness on $u$, transition from attractive low-angle to repulsive
high-angle grain boundaries, existence of overheated states -- 
are also present for the symmetric tilt grain boundaries of 
inclination $\phi=30^\circ$. However, for this inclination
there is no transition from single dislocations to dislocation 
pairs. This transition was also not observed in the 
three-dimensional PFC study with bcc symmetry in 
Ref.~\cite{Beretal08}. It can hence be concluded that its
occurrence depends on the detailed microscopic structure
of the grain boundary.

\begin{figure}
\begin{center}
\includegraphics[width = 7 cm]{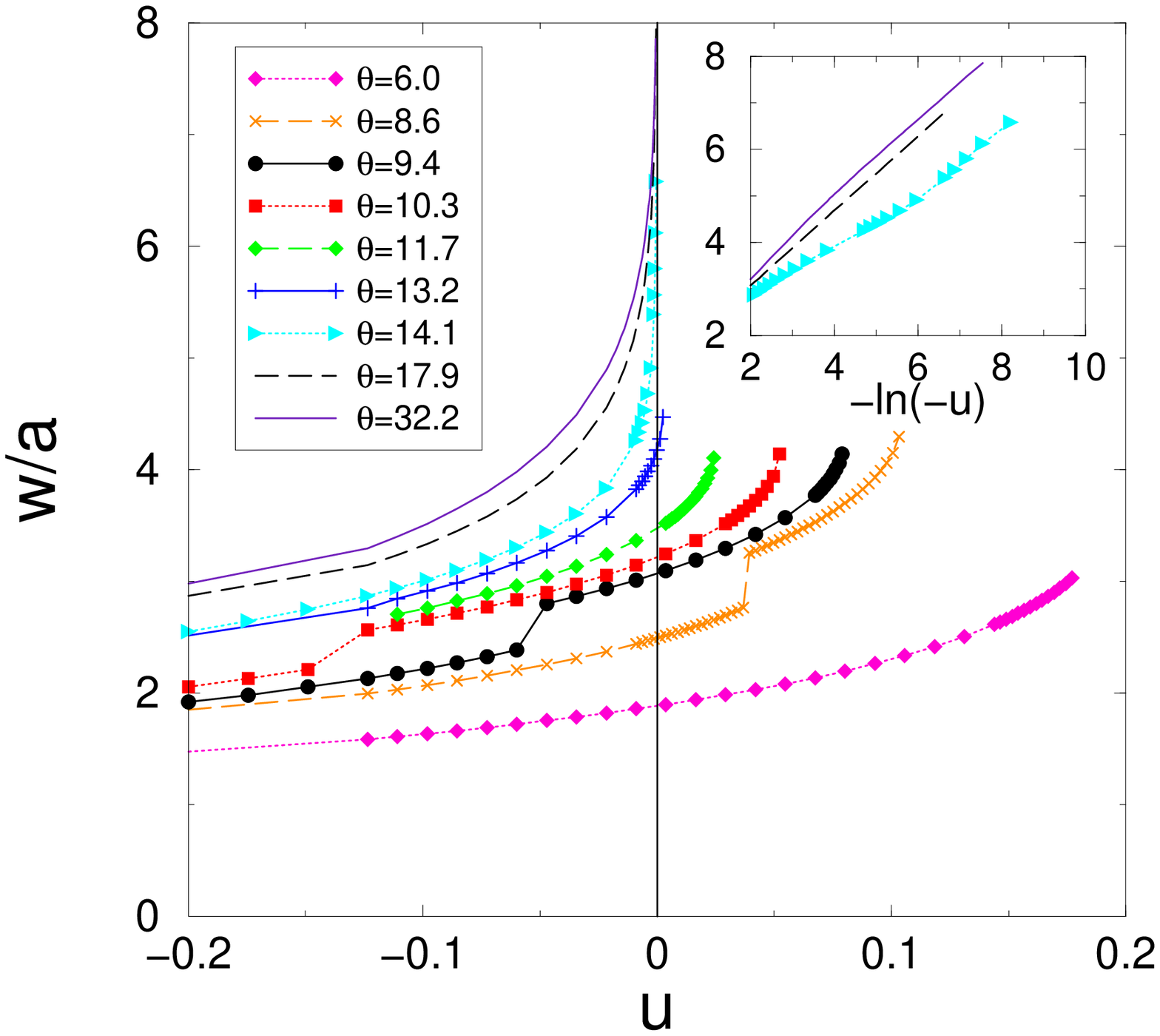}\\
\includegraphics[width = 7 cm]{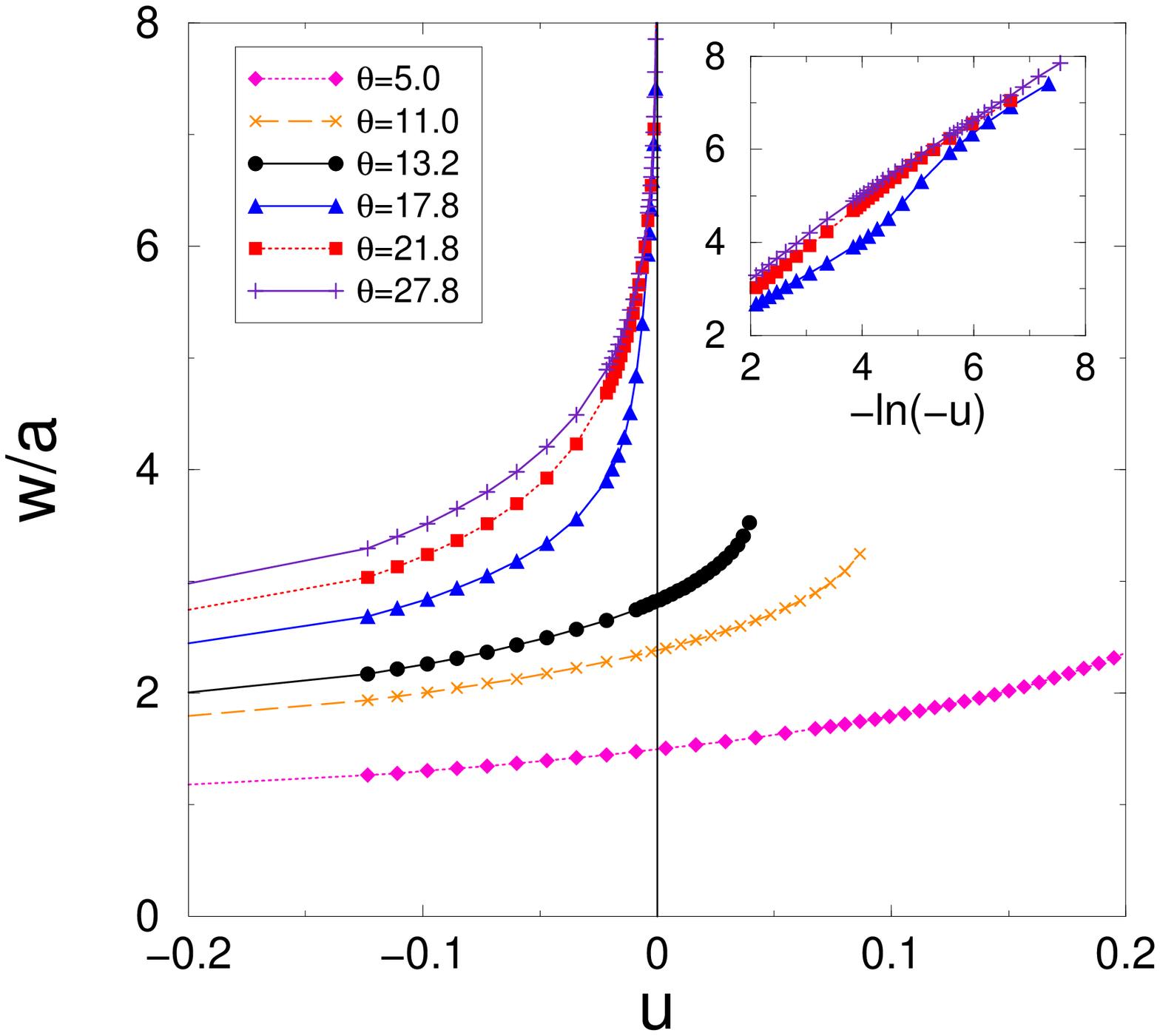}
\caption{(Color online)
Film thickness $w$ as a function of $u$ for various misorientations,
for symmetric tilt grain boundaries of inclination $\phi=0^\circ$
(top) and $\phi=30^\circ$ (bottom)
close to the melting point. All angles are given in degrees, and
a vertical line has been drawn at the melting point $u=0$. Inset:
Film thickness of the three largest angles versus $-\ln(-u)$:
the divergence of the film thickness is logarithmic.
\label{eps:Xdetail}}
\end{center}
\end{figure}

Let us now give a more detailed description of the transition
between the high-angle and the low-angle regimes. In 
Fig.~\ref{eps:Xdetail} we show the curves of $w$ versus $u$, 
for various misorientations, in the vicinity of the melting 
point, for the two inclinations $\phi=0^\circ$ and 
$\phi=30^\circ$, respectively. We recall (see the discussion
in Sec.~\ref{sec:numerics}) that due to the hexagonal
symmetry the two curves shown for $\phi=0^\circ$, 
$\theta=32.2^\circ$ and for $\phi=30^\circ$, $\theta=27.8^\circ$
actually describe the same grain boundary. All curves have been
calculated by simulations at fixed chemical potential. The
final state of a given run was used as initial condition for
the next one at slightly different chemical potential.

The insets of Fig.~\ref{eps:Xdetail} show the film thickness for the 
three largest misorientations for both inclinations, all corresponding 
to repulsive interfaces, versus $-\ln(-u)$. For large film 
thickness, the curves become linear, which is the dependence
that is expected for an exponential disjoining potential
from Eqs.~(\ref{eq:contactpotential}) and (\ref{wet}).
According to Eq.~(\ref{wet}), the slope of this linear part 
is the decay length $\delta$. We find a value of $\delta\approx 5.8$,
which is approximately half of the thickness of the solid-liquid 
interfaces $\delta_{\mathrm sl}\approx 12.5$, and comparable to the
wavelength of the dominant density waves of the hexagonal
structure (which is equal to $2\pi$ in our scaling).

It can be seen that the transition between repulsive and 
attractive behaviors occurs at an angle of 
$\theta_c\approx 14^\circ$ for both inclinations. This transition is 
smooth in the sense that the critical value $u^*$ where the 
solid bridges break decreases with increasing misorientation
and seems to tend to zero at the transition angle without 
exhibiting a jump. Furthermore, the thickness of the liquid
layer at the melting point, $w_m=w(0)$, increases with
misorientation and seems to diverge continuously when
$\theta_c$ is approached from below. The precise nature of 
this divergence remains undetermined. Its detailed study 
would require simulations in a narrow range of misorientations
close to the critical angle, which is quite cumbersome
because of the geometrical constraints that arise from the
finite size of the simulation box.

It is important to stress that this transition does {\em not}
coincide with a structural transition of the grain boundary.
The curves of $w$ versus $u$ for $\theta=32.2^\circ$ and 
$\theta=17.9^\circ$ in Fig.~\ref{eps:Xdetail} are very 
similar; however, the structure of these grain
boundaries is quite different. In all the snapshot pictures in
Fig.~\ref{eps:HighAngle},
the grain boundary is a plane of mirror symmetry for the
density field. This is not the case for $\theta=17.9^\circ$:
far from the melting point, this grain boundary consists
of individual dislocations such as the low-angle grain boundary
shown in Fig.~\ref{eps:LowAngle}.
The transition from single dislocations to dislocation pairs
also occurs, but far from the melting point, around $u=-1.4$.
When the melting point is approached, a continuous transition
from a state similar to the uppermost left picture in
Fig.~\ref{eps:LowAngle}
to one that looks like the uppermost left picture in
Fig.~\ref{eps:HighAngle}
occurs: the liquid pools around the dislocation pairs increase
in size and finally merge to give rise to a fairly homogeneous
film. The ``liquid pools'' separated by ``solid bridges'' are
therefore present in the vicinity of the melting point both
for repulsive and attractive grain boundaries.

In Fig.~\ref{eps:Xdetail}, jumps in the film thickness can
be seen in the curves for $\theta=8.6^\circ,\,9.4^\circ$,
and $\theta=10.3^\circ$; they correspond to the occurrence
of the transition from single dislocations to dislocation
pairs. The value of $u$ at which this transition occurs
increases with decreasing misorientation. As mentioned
before, for $\theta=17.9^\circ$, it occurs far below the
melting point; for $\theta=8.6^\circ$, it occurs only
above the melting point. For an even lower misorientation,
$\theta=6.0^\circ$, it does not occur at all: the liquid
pools around the single dislocations increase in size until
the solid ``bridges'' between them break.

\begin{figure}
\begin{center}
\includegraphics[width = 6 cm]{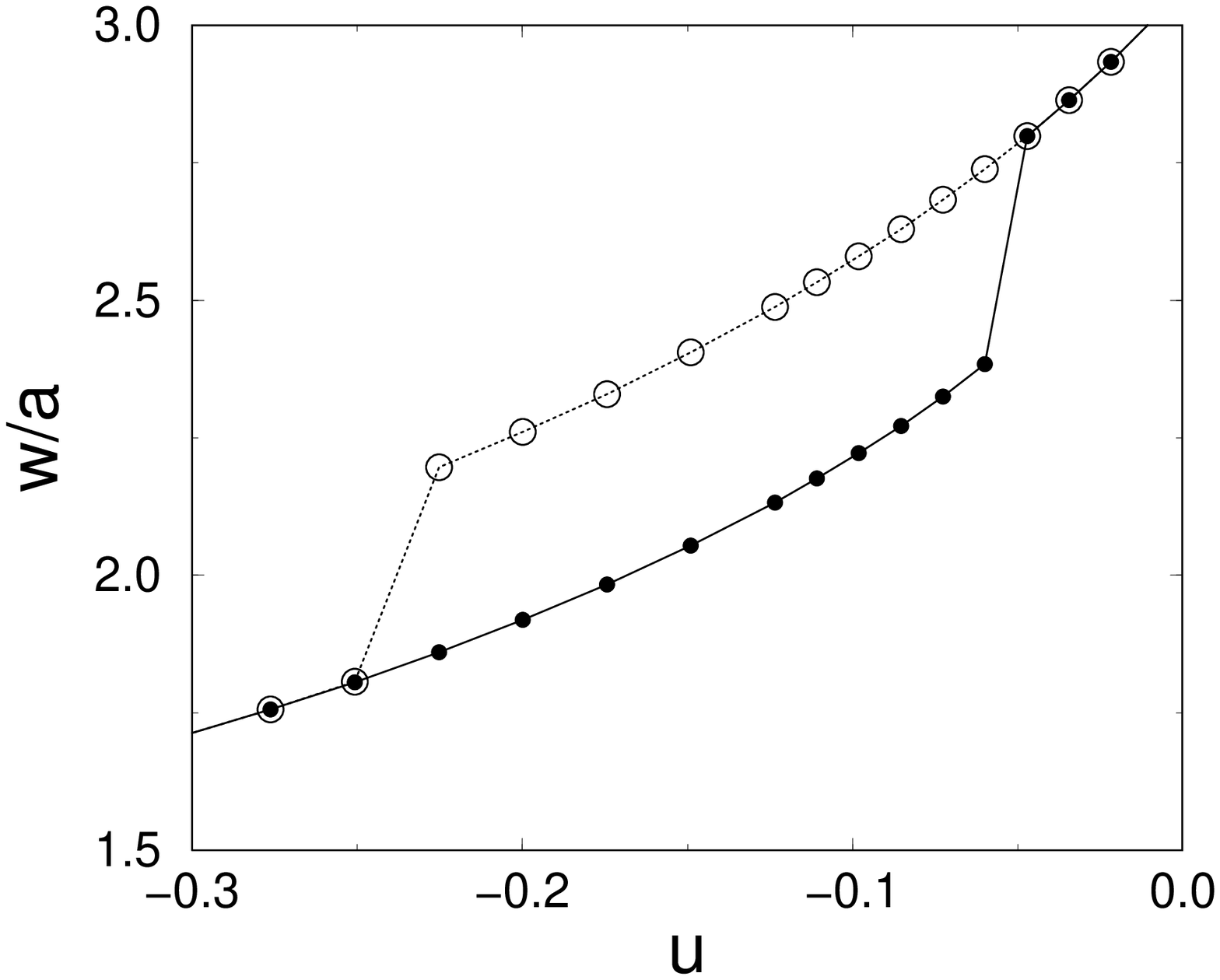} \\
\includegraphics[width = 5 cm]{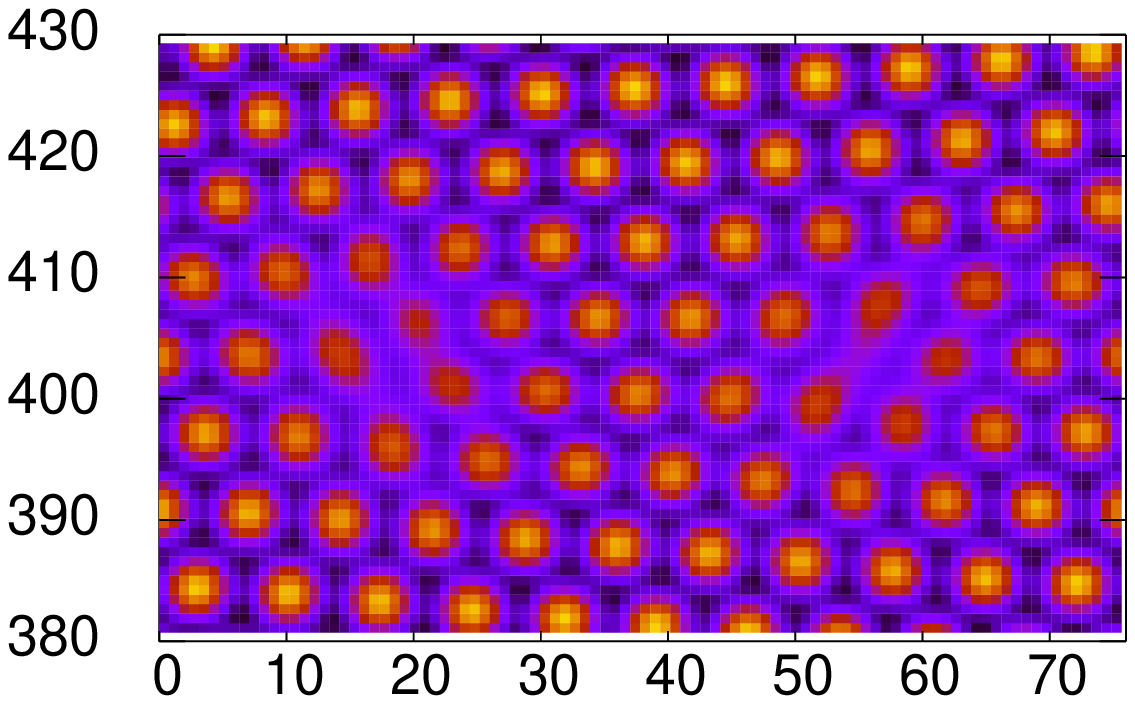} \\
\includegraphics[width = 5 cm]{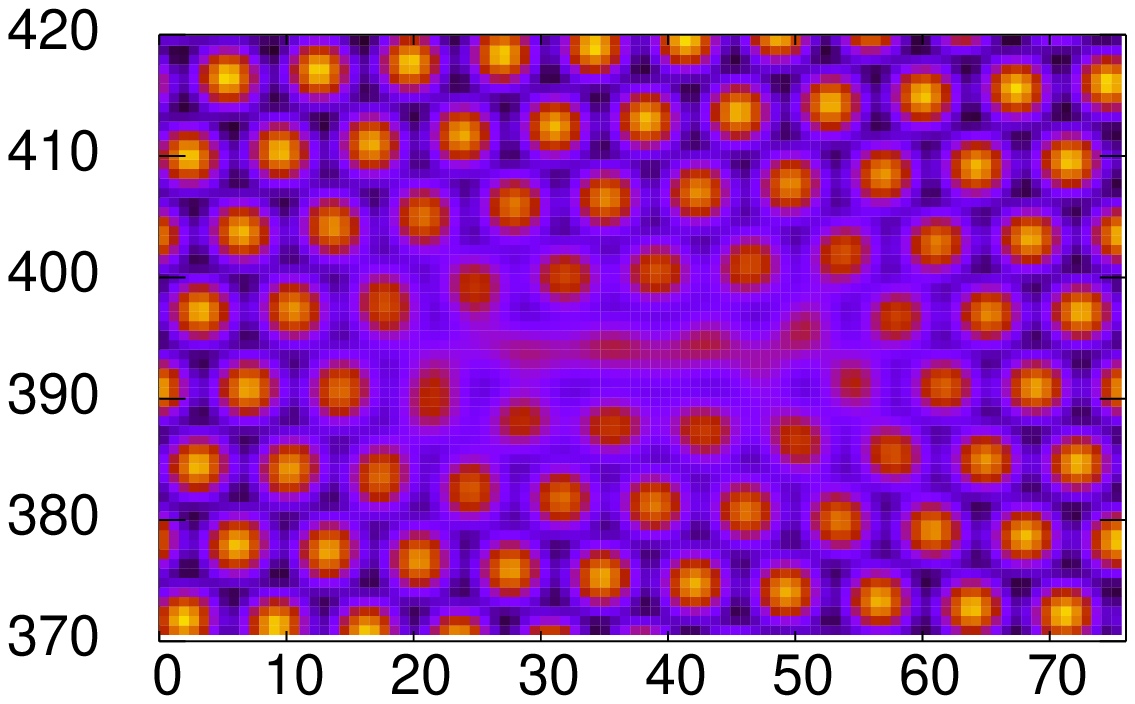} \\
\includegraphics[width = 5 cm]{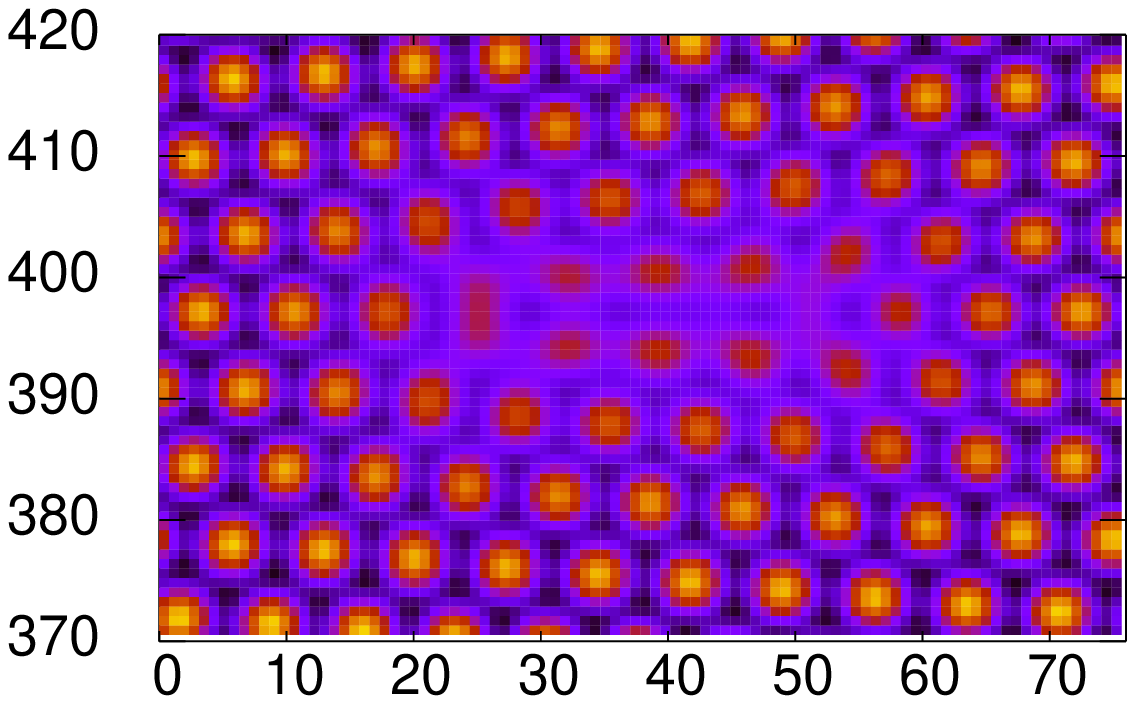}
\caption{
Various grain-boundary states for inclination $\phi=0^\circ$ 
and misorientation $\theta=9.4^\circ$.
The curve $w$ versus $u$ exhibits a hysteresis: filled symbols are
calculated with increasing $u$ (same curve as shown in 
Fig.~\protect\ref{eps:Xdetail}), and open symbols with decreasing $u$.
The snapshot pictures show different grain-boundary states, all at 
$u=-0.072$. Top: single-dislocation state, middle: dislocation pair
calculated with a single-dislocation state as initial condition,
bottom: dislocation pair calculated with the initial conditions
described in Sec.~\protect\ref{sec:numerics}.
\label{eps:pairtransition}}
\end{center}
\end{figure}

This structural transition exhibits a hysteresis. In 
Fig.~\ref{eps:pairtransition} we show two curves of $w$ versus
$u$ for $\theta=9.6^\circ$ that are computed in different ways. 
In the first, we start from a single-dislocation state at low 
values of $u$ and then perform successive simulations with
increasing $u$, taking the final state of the previous
simulation as initial condition. In the second, we start
from a dislocation-pair state and successively decrease
the value of $u$. It can be seen that there exists a range
of $u$ in which both single-dislocation and dislocation-pair 
states are stable. This indicates that there exist at least
two distinct branches of grain-boundary solutions. In addition, 
there exist different configurations for the dislocation-pair state, 
as is shown in the snapshot pictures in Fig.~\ref{eps:pairtransition}:
if the simulation is started from the initial conditions
described in Sec.~\ref{sec:numerics}, the dislocation
pair state exhibits a mirror symmetry with respect to the
plane of the grain boundary, as also seen in the snapshots
in Fig.~\ref{eps:LowAngle}.
The dislocation-pair states obtained starting from a 
single-dislocation state do not exhibit this symmetry. However, the 
size and shape of the ``liquid pool'' are quite similar, and the 
film thickness extracted from the two different configurations is 
identical up to the numerical precision. We have also found
that in the case of the low-angle grain boundary that does not 
exhibit the transition ($\theta=6.0^\circ$), dislocation-pair
states can be obtained starting from the initial condition in
Sec.~\ref{sec:numerics}; they form a second branch of solutions 
for this misorientation.

The hysteretic nature of this transition and the dependence of 
the final state on the initial conditions are clear indications
that the free-energy functional of the PFC model has several
distinct minima which correspond to grain-boundary states of
different grain-boundary energies. For low values of $u$, 
the single-dislocation states have a lower energy and
the dislocation-pair states correspond to a metastable
minimum, whereas the inverse is true for high values of $u$.
Since the extraction of the grain-boundary energy is delicate,
we have not pinpointed the exact value of $u$ where the two 
states have equal energies. But from the general phenomenology
of hysteretic transitions, it can be expected to lie approximately 
in the middle of the bistable range. Furthermore, since our
simulations do not include thermal fluctuations, the termination
of the metastable branches corresponds to the disappearance of the
local metastable minimum.

The existence of metastable states raises the question of whether
other grain-boundary configurations, distinct from the ones
depicted in Figs.~\ref{eps:HighAngle} to \ref{eps:pairtransition}, 
might exist. We investigated this question by performing
several runs with different initial conditions for numerous
parameter sets. We did not find any new grain-boundary states
in the vicinity of the melting point. For states far from the 
melting point, we have occasionally observed distinct configurations
that exhibit differences in the local arrangements of the ``atoms'' 
around the dislocations and different total number of ``atoms'',
which is possible since, even at fixed total mass, the total
number of ``atoms'' is not fixed in the PFC model. No further 
investigation of these multiple states was carried out.

The curves of $w$ versus $u$ for intermediate misorientations
that consist of dislocation pairs exhibit a vertical slope at 
the ``break point''. It is tempting to believe that the solution 
branch continues beyond that point and bends back to reach $u=0$ 
when $w\to\infty$. This would be expected if the state of the
grain boundary can be properly described by the single 
variable $w$; such solutions have been found in phase-field 
studies of grain-boundary premelting \cite{Lobkowski02,Tang06}.
In our PFC model, these solutions cannot be obtained in
simulations at constant chemical potential, since they are 
unstable. We have tried to obtain these states by simulations 
with fixed total mass. In this case, mass conservation yields 
a constraint on the film thickness which should stabilize these 
states. However, our attempts were not successful due to the 
occurrence of a new instability. Since there are always 
two distinct grain boundaries in our system due to the
periodic boundary conditions, a symmetry breaking can 
occur which leads to the formation of a ``thick'' and a 
``thin'' liquid film instead of two liquid films
of equal thickness. This indeed happens when the film 
thickness is larger than the value corresponding to the 
``turning point''. A simple explanation for this
instability will be given below. 
In contrast, all the curves $w(u)$ for low-angle grain boundaries 
with inclination $\phi=30^\circ$ as well as the curve 
for the lowest misorientation for $\phi=0^\circ$ ($\theta=6.0^\circ$)
do not exhibit a turning point, but break off with a finite 
slope. These grain-boundary states all consist of single dislocations. 

From these results it can be concluded 
that the mechanisms that lead to the breaking of the 
``solid bridges'' and to the instability of the overheated 
solution branches depend on the detailed structure of the 
grain boundary. Qualitatively, the difference in behavior can
be understood from geometric considerations. As mentioned above,
a vertical slope at the break point would be expected for
homogeneous liquid films that can be faithfully described
by a single variable, the film thickness $w$. The elongated 
liquid pools surrounding the dislocation pairs are more similar 
to a homogeneous liquid film than the round liquid pools 
surrounding single dislocations. Thus, it is not surprising
that the behavior of the former is closer to the one of a
homogeneous film.

It is clear from the above results that the
description of a grain boundary by a single variable
(the thickness) is very crude. However, we have found no
other obvious quantity that could play the role of a 
supplementary state variable. Therefore, for all the 
following developments we will restrict our level of 
description to the single variable $w$, leaving a more 
detailed investigation as a subject for further study.
Also note that, in principle, a distinct grain-boundary 
energy and disjoining potential are associated with each
of the distinct solution branches. To simplify the
picture, we will ignore this fact and display in the
following unique curves for the grain-boundary energy
and the disjoining potential. Since the film thickness
$w$ exhibits a jump, $V(w)$ has a ``step'', and the
grain-boundary energy a discontinuity in slope. These
features are, however, so small that they can be
hardly distinguished in the following plots.

\subsection{Grain-boundary energy}

\begin{figure}
\begin{center}
\includegraphics[width = 8 cm]{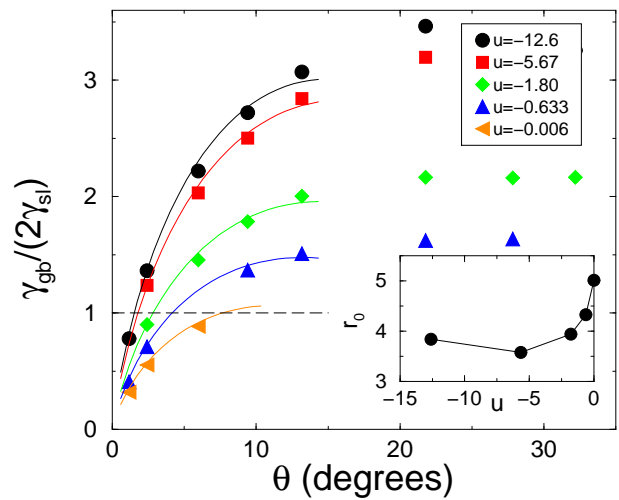}
\caption{(Color online)
Ratio of grain-boundary energy to twice the solid-liquid free
energy for grain boundaries of inclination $\phi=0^\circ$ 
as a function of the misorientation $\theta$, shown for 
different chemical potentials $u$. The lowest curve, $u = -0.006$,
is very close to the melting point, while the upper curve is 
far inside the solid region. The lines are fits to the Read-Shockley 
law, Eq.~(\ref{Eq:Pfc:RS2}), using the values of $\gamma_{\mathrm gb}$ 
for $\theta<15^\circ$. The shear modulus has been fixed to the 
theoretical value at the corresponding chemical potential, 
and the only fit parameter is the dislocation core radius $r_0$.
Inset: The core radius $r_0$ obtained from the fits as a function
of chemical potential. For comparison, the lattice constant 
is $a\approx 7.255$ and the value estimated by Elder and 
Grant~\cite{Eldetal04} is $r_0\approx 4.4$.
\label{eps:GBE-Summary}}
\end{center}
\end{figure}

We calculate the grain-boundary energy as described in 
Sec.~\ref{SubSec:Numerics:ContactPontential} by performing
simulations at fixed density $\bar\psi$ for several different system
sizes $L_y$ and using Eq.~(\ref{eq:fvsLy}) to extract $\gamma_{\mathrm gb}$.
In order to keep the presentation 
consistent, we will nevertheless discuss the results as a function 
of $u$, which can be easily obtained for given density using the 
curve $\mu_s(\bar\psi)$. All the data shown in this
subsection are for grain boundaries of inclination $\phi=0^\circ$.
The grain-boundary energy is plotted versus misorientation
in Fig.~\ref{eps:GBE-Summary}
for various values of the chemical potential. Two clear 
tendencies can be seen. First, for any fixed misorientation, 
$\gamma_{\mathrm gb}$ increases monotonously when $u$ decreases. Second,
for a fixed supersaturation, $\gamma_{\mathrm gb}$ increases monotonously
with the misorientation for small angles.

The latter dependency can be well understood in terms of the 
Read-Shockley law \cite{ReadShockley}, 
\begin{equation}
\gamma_{\mathrm gb} = \frac {G a }{4\pi \alpha (1-\sigma)} 
\theta \left[ 1 - \ln (2\pi\theta) + \ln(\alpha a/r_0)\right] \label{Eq:Pfc:RS2} \,, 
\end{equation}
where $r_0$ is the core radius of the dislocations, $a$ is the lattice 
constant, $\alpha=\sqrt{3}/2$ is the ratio of the distance between
close-packed planes and the lattice spacing, $G$ is the shear modulus, 
and $\sigma$ is Poisson's ratio.
The elastic properties of the PFC model can be determined analytically 
in the one-mode approximation~\cite{Eldetal02,Eldetal04}.
The resulting elastic constants are $C_{11}/3 = C_{12} = C_{44} = 
\left(3\bar\psi - \sqrt{15 \epsilon - 36 \bar\psi^2}\right)^2/75$. 
The bulk modulus can then be calculated to be $Y=2C_{44}$ and 
the shear modulus $G=C_{44}$. Furthermore, the three-dimensional 
Poisson's ratio is $\sigma = (3Y-2G)/[2(3Y+2G)] = 1/4$.
An estimation for the core radius, $r_0 = a\exp(-0.5) \approx 4.4$
has also been given~\cite{Eldetal04}. Note that we have used here the 
standard (three-dimensional) version of the Read-Shockley law and
the elastic constants. Tt can be shown \cite{remark3} that this is identical 
to the two-dimensional expressions given by Elder and Grant \cite{Eldetal04}.
Furthermore, this formula differs from the standard one for cubic
materials by the presence of the factors $\alpha$, which are due to
the fact that the average distance $d$ between dislocations is 
$d\sim \alpha a/\sin\theta$ (instead of $d\sim a/\sin\theta$
for a cubic material). It should also be emphasized that this 
expression is only valid for grain boundaries of inclination 
$\phi=0^\circ$; in the general case, several additional terms 
depending on the inclination angle have to be included \cite{ReadShockley}.

For each supersaturation, we fixed the shear modulus to
its density-dependent analytical value, and performed
a least-squares fit of our data to the Read-Shockley law, 
with $r_0$ as the only fit parameter. Since Eq.~(\ref{Eq:Pfc:RS2})
is only valid for small misorientations, for the fit only 
systems with $\theta<15^\circ$ have been included. It can
be seen that the fit is excellent. In the inset, the core 
radius $r_0$ obtained from the fit is shown as a function 
of $u$. It is almost constant and close to the theoretically 
estimated value for large values of $|u|$, and increases when 
the coexistence region is approached ($u\to 0$).

\begin{figure}
\begin{center}
\includegraphics[width = 8 cm]{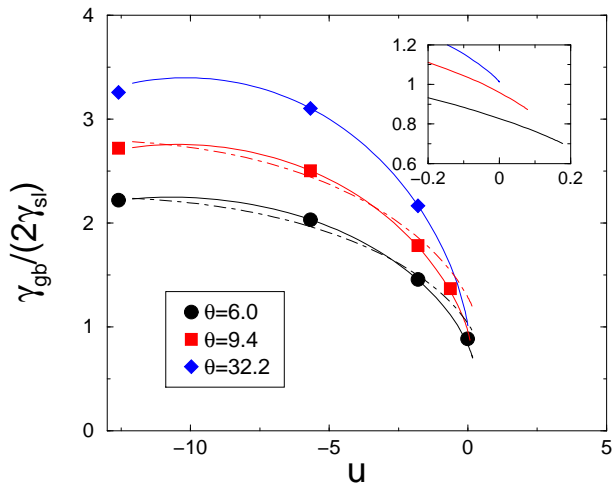}
\caption{(Color online) Symbols: grain-boundary energy determined 
from direct simulations, lines: grain-boundary energy calculated
by thermodynamic integration, dash-dotted lines: prediction 
of the Read-Shockley law with $r_0=3.75$ and elastic constants 
depending on $u$. See text for details.
\label{eps:ggbvspsi}}
\end{center}
\end{figure}

It turns out that the variation of the grain-boundary energy
with $u$ will be crucial for the further discussions. We
recall that this variation is directly linked to the liquid
film thickness by Eqs.~(\ref{eq:thermintegration}) and 
(\ref{eq:filmthickness}). The grain-boundary energy is shown as a 
function of $u$ for three selected misorientations in 
Fig.~\ref{eps:ggbvspsi}. The symbols are values that have
been directly obtained from simulations with varying system
size. The full lines are obtained by integrating
Eq.~(\ref{eq:thermintegration}), where the integration 
was started from the data point at $u=-5.674$.
It is clear that the two procedures give fully consistent
results. In the inset, the data for $\gamma_{\mathrm gb}$ obtained by 
integration are shown in the vicinity of the melting point.
It should be mentioned that direct calculations of the 
grain-boundary energy in this regime are quite difficult, since
the grand potential differences between solid and liquid 
(and hence the driving forces) are small, so that long
equilibration times are needed.
The different behaviors of repulsive and attractive 
grain boundaries can be clearly seen. For the repulsive
grain boundary (upper curve), $\gamma_{\mathrm gb}$ tends to
$2\gamma_{\mathrm sl}$ from above. Since $\psi_{\mathrm{exc}}\sim\ln(-u)$
close to the melting point, we have 
$\gamma_{\mathrm gb} - 2\gamma_{\mathrm sl} \sim u\ln(-u)+u$, as expected
from Eq.~(\ref{gben}) of the sharp-interface theory. The curve 
has an infinite slope at $u=0$ (corresponding to a diverging film
thickness), but the logarithmic divergence is too slow to
be clearly distinguished in the figure. For the attractive
grain boundaries (the lower two curves), $\gamma_{\mathrm gb}$
becomes lower than $2\gamma_{\mathrm sl}$ {\em before} the melting
point is reached. It continues to decrease beyond the
melting point, until the metastable solution branch ends.

As mentioned above, the dislocation core radius extracted
from the fits to the Read-Shockley law is almost constant
over a wide range of $u$. An interesting corollary of this
finding is that the variation of the grain-boundary energy 
with chemical potential can be accounted for almost entirely 
by the change in the elastic constants. To illustrate this point, 
we show in Fig.~\ref{eps:ggbvspsi} as dash-dotted lines
the predictions of the Read-Shockley law with a constant
value for the core radius $r_0=3.75$ for the two lowest 
misorientations. Clearly, the main variation of
$\gamma_{\mathrm gb}$ with supersaturation is well reproduced.
The ratio of the predicted to the numerical value remains
close to unity up to $u\approx -2$ and then increases
sharply to about $1.4$ at the melting point. This is natural 
since the assumption of constant core radius breaks down.
The highest misorientation shown in Fig.~\ref{eps:ggbvspsi} is 
too large for the Read-Shockley law to be applicable. But from the 
figure it is clear that the variation of $\gamma_{\mathrm gb}$ with the
chemical potential is very similar to the one of the low-angle
grain boundaries. It is therefore reasonable to assume that
this variation is also mainly controlled by the elastic constants.

\subsection{Disjoining potential}

\begin{figure}
\begin{center}
\includegraphics[width = 8 cm]{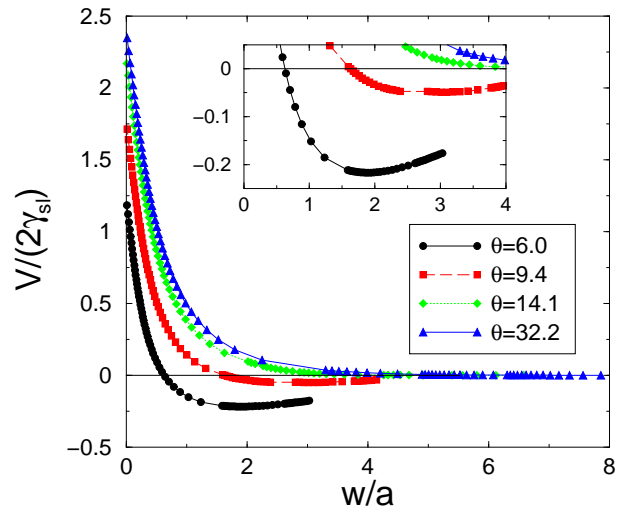}
\caption{(Color online) Disjoining potential for inclination 
$\phi=0^\circ$ and various misorientations (inset: detailed
view of the region around the minimum). The angles are given 
in degrees.
\label{eps:Potential}}
\end{center}
\end{figure}

As outlined in Sec.~\ref{SubSec:Numerics:ContactPontential}, two methods 
can be used to extract the disjoining potential from the simulation data: 
$V'(w)$ can be integrated using the data of $w(\mu)$, or $V(w)$ can be
directly deduced from the grain-boundary energy using 
Eq.~(\ref{eq:Vggbrelation}). Both methods 
yield consistent results that are shown in Fig.~\ref{eps:Potential}.
For the high-angle grain boundaries, $V(w)$ decreases
monotonously; it can be actually quite well described by an 
exponential function as in Eq.~(\ref{eq:contactpotential}).
In contrast, for the low-angle grain boundaries, the disjoining 
potential is non-monotonous: starting from a positive value at
$w=0$, it decreases, falls below zero and exhibits a 
minimum for some intermediate values of $w$. It then
starts to increase until it reaches the point where
the curve $w(\mu)$ terminates.

\section{Discussion}
\label{sec:Discussion}

It is instructive to discuss some aspects of our above results
in more detail and to compare them with the predictions of
the sharp-interface theory. Three questions are of particular
interest: what is the interpretation of the non-monotonous 
disjoining potentials, what determines the critical angle for 
the attractive-to-repulsive transition, and what can be
said about the overheated grain boundaries and about
transitions between different grain-boundary states ?

The shape of the disjoining potential for low-angle grain
boundaries can obviously not be described by the simple
exponential form of Eq.~(\ref{eq:contactpotential}). This
potential corresponds to a short-range repulsion, but a long-range 
attraction. We have not found a simple analytical formula for 
this potential; however, a few of its properties can be
readily understood. For instance, Eq.~(\ref{eq:pdef})
tells us that $V'(w)=0$ implies $\omega_s=\omega_l$,
which is only the case for $\mu=\mu_{\mathrm{eq}}$ ($u=0$). The
minimum of the curve $V(w)$ corresponds therefore to
the intersection of the curves $w(u)$ with the $u=0$ 
axis in Fig.~\ref{eps:Xdetail}. Since Eq.~(\ref{eq:Vggbrelation}) 
yields, for $\omega_s=\omega_l$,
$V(w)=\gamma_{\mathrm gb}-2\gamma_{\mathrm sl}$, this implies that the depth
of the potential well is given by the difference of the
grain-boundary energy {\em at the melting point} and
twice the solid-liquid free energy. Furthermore, 
the value $V(0)$ corresponds to the grain-boundary 
energy of a completely dry grain boundary $\gamma_{\mathrm gb}^0$.
The height of the ``repulsive part'' of the disjoining potential 
is therefore given by the difference between this value and
the grain-boundary energy at the melting point. Any system in 
which the grain-boundary energy increases with decreasing
homologous temperature will therefore exhibit a repulsive part
in the disjoining potential, even if the grain-boundary
is attractive at the melting point. In addition, from
Eq.~(\ref{eq:thermintegration}) and Eq.~(\ref{eq:eqfilmthickness})
it is easily seen that the variation of the grain-boundary energy
with $u$ is proportional to the negative of the liquid film thickness.
Therefore, the disjoining potential is repulsive below the
melting point for {\em any} grain boundary that exhibits a 
finite film thickness.

It is also easy to show that the points where the curve $w(\mu)$ 
exhibits a vertical tangent correspond to an inflection point 
in the potential $V(w)$. For this, it is sufficient to take
the derivative with respect to $w$ of Eq.~(\ref{eq:eqfilmthickness}),
which yields
\begin{eqnarray}
V''(w) & = & \left(\frac{\partial\omega_s}{\partial\mu}-
                   \frac{\partial\omega_l}{\partial\mu}\right)\frac{d \mu}{d w} \nonumber \\
       & = & \left(\bar\psi_l-\bar\psi_s\right)\frac{d \mu}{d w}.
\end{eqnarray}
The sign of the second derivative of the disjoining potential is
hence determined by the derivative $d\mu/d w$, which is zero
at the turning point of $w(\mu)$. As a consequence, $V(w)$ is
concave for large values of $w$. This yields a simple explanation
for the instability that leads to a symmetry breaking between the 
two grain boundaries in the simulation box, which was described
in Sec.~\ref{SubSection:ResultsStructure}: at fixed density, the sum 
of the two film thicknesses $w_1$ and $w_2$ is approximately fixed by
the lever rule. If $w_1=w_2=w$, and $w$ is located in the concave 
part of the potential, the system can lower its total energy by 
making one film wider and the other one thinner.

Let us now consider the transition from attractive to
repulsive grain boundaries. As discussed in Sec.~I,
the sharp-interface theory predicts this transition to occur
when $\gamma_{\mathrm gb}=2\gamma_{\mathrm sl}$. Now consider the different 
curves of $\gamma_{\mathrm gb}$ versus misorientation shown in 
Fig.~\ref{eps:GBE-Summary}.
For the two lowest values of $u$, this curve intersects the 
line corresponding to twice $\gamma_{\mathrm sl}$ for a misorientation of
$\theta\approx 2^\circ$, much smaller than the transition angle 
obtained from the curves $w(u)$ in Fig.~\ref{eps:Xdetail}. However,
with increasing $u$, this intersection point
moves toward larger angles, and for the highest value of $u$
investigated, the intersection is at about $9^\circ$.
Furthermore, it was shown above that when the disjoining potential
exhibits a minimum, its value at this minimum is equal to 
$\gamma_{\mathrm gb}-2\gamma_{\mathrm sl}$ at the melting point. Since the
transition to repulsive grain boundaries occurs when this
minimum disappears, it is to be expected that the correct 
transition angle is obtained when the criterion $\gamma_{\mathrm gb}=2\gamma_{\mathrm sl}$
is used with the grain-boundary energy calculated exactly at 
the melting point. 

\begin{figure}
\begin{center}
\includegraphics[width = 8 cm]{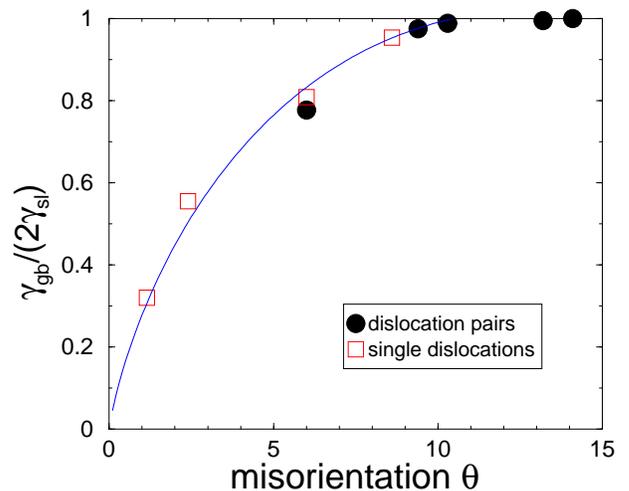}
\caption{Ratio of the grain-boundary energy and the melting point
$\gamma_{g}^m$ and twice the solid-liquid free energy as a function 
of misorientation for inclination $\phi=0^\circ$. The line is a 
fit to the Read-Shockley law for the points with $\theta<10^\circ$.
\label{eps:gbe_melting}}
\end{center}
\end{figure}

In order to obtain more precise information on this question,
it would be desirable to have accurate values for the 
grain-boundary energy at the melting point as a function of misorientation.
However, as already mentioned, it is numerically very difficult to 
obtain values for the grain-boundary energy close to the melting
point, especially for angles close to the repulsive-to-attractive
transition, since the grand potential differences between the
different states become extremely small. The best way to obtain
reliable data close to the transition is to integrate 
Eq.~(\ref{eq:thermintegration}) up to the melting point. 
The result is shown in Fig.~\ref{eps:gbe_melting}; we estimate
the error bars for these data to be on the order of the size
of the symbols. For low-angle grain boundaries, the dependency
of $\gamma_{\mathrm gb}$ on $\theta$ can still be described by the
Read-Shockley law. However, for higher angles, when the
grain boundaries consist of dislocation pairs, the dependence
of $\gamma_{\mathrm gb}$ on $\theta$ is extremely weak: for 
$\theta=10.4^\circ$, which is the first point that clearly
deviates from the Read-Shockley law, 
$\gamma_{\mathrm gb}/(2\gamma_{\mathrm sl})\approx 0.99$, so that the 
variation of $\gamma_{\mathrm gb}$ between this misorientation and the
first repulsive grain boundary at $14.1^\circ$ is only about $1$\%. 
Clearly, it is very difficult to describe precisely this regime.
In addition, it is not clear whether it is generic, since the 
grain boundaries of inclination $\phi=30^\circ$ do not exhibit 
the structural transition to dislocation pairs.

A very interesting point is that the Read-Shockley law is
still valid for low-angle grain boundaries, even at the
melting point. This can be used to obtain a reasonable
estimate for the critical angle as the solution of
the equation
\begin{equation}
\frac {G a }{4\pi \alpha (1-\sigma)} \theta_c 
\left[ 1 - \ln (2\pi\theta_c) + \ln(\alpha a/r_0)\right] = 
2 \gamma_{\mathrm sl}.\label{modwetfin}
\end{equation}
However, it is crucial to take into account the variation of the 
grain-boundary energy with chemical potential (or temperature).
Indeed, the values for the grain-boundary energy are
usually determined in experiments or atomistic simulations 
for temperatures far below the melting point. As pointed out
above, if these values are used to predict the critical 
angle, a completely wrong result is obtained. The variation
of the grain-boundary energy with chemical potential (or
temperature) arises from two distinct effects: the variation
of the shear modulus and the premelting around dislocations,
which leads to an increase of the core radius in the 
Read-Shockley law as shown in Fig.~\ref{eps:GBE-Summary}. 
If the ``low-temperature'' values for both
$G$ and $r_0$ are used in Eq.~(\ref{modwetfin}), we find
$\theta_c\approx 2^\circ$ (see Fig.~\ref{eps:GBE-Summary}), 
clearly too low. If only the variation of $G$ is included
(that is, Eq.~(\ref{modwetfin}) is used with the value of
the shear modulus at the melting point, but with the
low-temperature value $r_0=3.5$ for the core radius), 
the prediction becomes $\theta_c\approx 6^\circ$.
Finally, the improved estimate corresponding to
Eq. (\ref{modwet2}) is obtained when the values at the melting 
point of both the shear modulus and the core radius are used,
which yields the prediction $\theta_c\approx 10^\circ$. 
The curve of $\gamma_{\mathrm gb}^m$ that is obtained with these
parameters is shown in Fig.~\ref{eps:gbe_melting}. Clearly,
the limited accuracy of this final estimate is due to 
the fact that the Read-Shockley law applies to grain-boundary 
states with individual dislocations. Thus, it does not take take 
into account the complex structural changes in the boundary,
which will tend to reduce the grain-boundary energy further from its
value estimated from the Read-Shockley law. Therefore, the value 
of $\theta_c$ obtained from this law is likely to be a lower bound 
estimate of the actual value.

\begin{figure}
\begin{center}
\includegraphics[width = 8 cm]{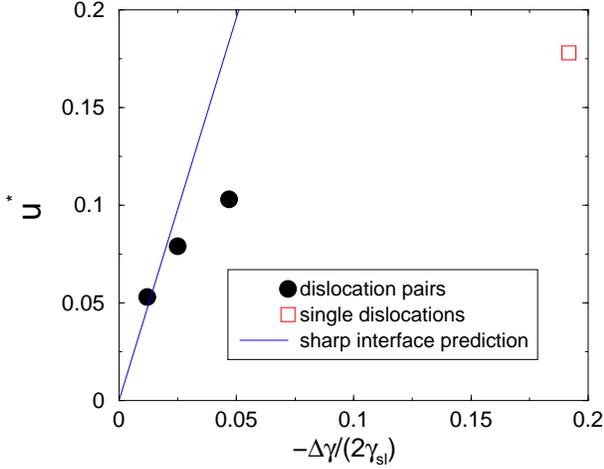}
\caption{Critical value $u^*$ that corresponds to the limit of 
superheated states associated with the breaking of solid bridges.
Symbols: simulation results, line: sharp-interface prediction
according to Eq.~(\protect\ref{eq:ustar}).
\label{eps:overheat}}
\end{center}
\end{figure}

Let us now come to the estimation of the critical value $u_*$
that limits the range of ``overheated'' states. The transposition
of the sharp-interface prediction, Eq.~(\ref{superheat}), to our
variables is
\begin{equation}
\omega_l(u^*)-\omega_s(u^*)=-\frac{\Delta \gamma}{\delta}.
\end{equation}
Expanding the grand potential around the melting point, and
using the definition of $u$, Eq.~(\ref{eq:udef}), we find
\begin{equation}
u^* = -\left[\frac{2\gamma_{\mathrm sl}}
     {\left.\frac{\partial\mu}{\partial\bar\psi_s}\right|_{\bar\psi_s^{\mathrm{eq}}}
       \left(\bar\psi_s^{{\mathrm{eq}}}-\bar\psi_l^{{\mathrm{eq}}}\right)^2 \delta}\right] 
     \frac{\Delta\gamma}{2\gamma_{\mathrm sl}}.
\label{eq:ustar}
\end{equation}
The ratio in brackets on the right-hand side can be calculated using
the values from Table~\ref{table1} and the values $\delta\approx 5.8$
and $2\gamma_{\mathrm sl}\approx 0.00192$
extracted from our simulations (see Fig.~\ref{eps:Xdetail}).
From the preceding discussion, it is clear that a reasonable
estimate can only be obtained if $\Delta \gamma$ is calculated
with the grain-boundary energy at the melting point. 
In Fig.~\ref{eps:overheat}, we plot the values of $u^*$
obtained from our simulations (that is, the values of $u$
where the curves of $w(u)$ terminate) versus 
$-\Delta\gamma/(2\gamma_{\mathrm sl})$, using the values for $\gamma_{\mathrm gb}$
in Fig.~\ref{eps:gbe_melting}, together with the theoretical
prediction. It can be seen that this prediction gives reasonable
values for the grain boundaries close to the transition that
consist of dislocation pairs, even if the simulation data cannot
be well described by a straight line. In contrast, the sharp-interface
theory strongly overestimates the value of $u^*$ for
low-angle grain boundaries consisting of single dislocations.

\begin{figure}
\begin{center}
\includegraphics[width = 8 cm]{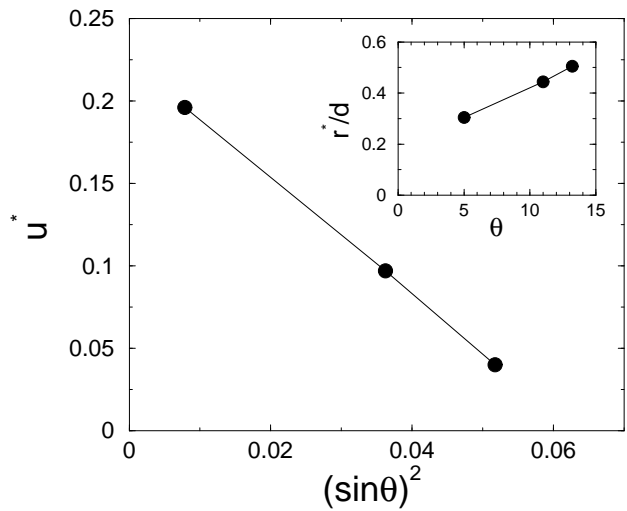}
\caption{Critical value $u^*$ versus $\sin^2\theta$ for
low-angle grain boundaries of inclination $\phi=30^\circ$.
Inset: Ratio of the liquid-pool radius $r^*$ where the break-off
occurs and the dislocation spacing $d$ versus misorientation.
\label{eps:ustarplot}}
\end{center}
\end{figure}

The failure of the sharp-interface theory to predict the superheated
range of grain boundaries is not surprising since the liquid
phase domains consist of liquid pools instead of a thin liquid
film of constant thickness as assumed in this theory. Recently, 
Berry {\it et al.} \cite{Beretal08} developed a simple theory
of grain-boundary wetting tailored to the liquid-pool geometry, 
which assumes that wetting occurs when pools coalesce, or, equivalently,
when their radius $r$ is equal to half of the distance $d$ between 
dislocations. By calculating the shift of the melting point due to 
the dislocation elastic strain energy, they also obtained the scaling
relation $u^*\sim -(a/d)^2$, where the dimensionless proportionality 
constant is related to the elastic constants. This scaling relation, 
together with the geometrical coalescence condition
$r=d/2=a/(2\sin\theta)$, yields the prediction $u^*\sim -\sin^2\theta$.
As shown in Fig.~\ref{eps:ustarplot}, our results for the
grain boundaries of the $\phi=30^\circ$ inclination that consist
of unpaired liquid pools indeed show that $u^*$ is reduced by an amount 
proportional to $\sin^2\theta$, consistent with this prediction. One 
important difference, however, is that the $\theta=0$ intercept of the 
curve $u^*(\theta)$ is finite in our simulations, consistent with 
the existence of superheated metastable grain boundaries, while 
the theory of Berry {\it et al.} \cite{Beretal08} predicts
that liquid pools always coalesce below the melting point ($u^*(0)=0$).
We expect $u^*(\theta)$ to be generally positive in the limit of
vanishing misorientation since a finite bulk thermodynamic
driving force favoring the liquid phase is necessary
to overcome the nucleation barrier imposed by the solid-liquid 
interfacial energy, which remains finite even in the presence of 
elastic strain energy around the dislocation cores.

It is interesting to note that the linear interpolation 
of the data in Fig.~\ref{eps:ustarplot}
predicts that $u^*$ vanishes for $\sin^2\theta\approx 0.06$,
which corresponds to a misorientation of $14.6^\circ$, in good
agreement with our previous estimate for $\theta_c$. For misorientations
above this value, no overheated states can exist, and there is hence
no discontinuous transition between ``dry'' and ``wet'' 
grain-boundary states.

To further test this theoretical picture for the
$\phi=30^\circ$ inclination, we have extracted the
pool radius from the data for the film thickness using
the simple geometrical transformation
\begin{equation}
r = \sqrt{\frac{wL_x}{n_d\pi}},
\label{eq:rdef}
\end{equation}
where $n_d$ is the number of dislocations present in the system.
Note that this pool radius differs from the dislocation core 
radius $r_0$ extracted from the fits to the Read-Shockley law. 
As shown in the inset of Fig.~\ref{eps:ustarplot}, the pool radius at
the break-off point, $r^*$, is not proportional to the dislocation
spacing. Furthermore, the size of the liquid pools, as defined by
Eq.~(\ref{eq:rdef}) (which is of course equivalent to a Gibbs
construction performed for a cylinder around a dislocation instead
of a flat homogeneous liquid layer), is not uniquely determined by
$u$, but also depends on the misorientation. This
indicates that the above picture needs to be refined in order to
better understand the condition for the coalescence of
liquid pools above the melting point. 

As a last point, it should be recalled that diffuse-interface 
theories of grain boundaries where the grain orientation 
is treated as a scalar order parameter have shown the possibility 
that two distinct grain-boundary states of markedly different widths 
can exist at the same temperature \cite{Lobkowski02,Tang06}. In 
contrast, aside from the dislocation-pairing hysteretic transition,
we have found here the grain-boundary width to be
uniquely determined at fixed chemical potential. However, we 
cannot rule out the existence of such two-state coexistence 
for crystal structures and grain-boundary orientations other
than those investigated here, or in a narrow range of chemical 
potential very close to the melting point where numerical calculations
with the PFC model become exceedingly difficult. Clearly,
this question warrants further investigation.

\section{Conclusions and outlook}

We have performed a detailed study of grain-boundary
premelting using the phase-field crystal model. Our results
demonstrate that there is a qualitative difference
between high-angle ``repulsive'' and low-angle ``attractive''
grain boundaries. In the former, a continuous liquid film
forms below the melting point, exhibiting a width that
diverges when the melting point is approached from below.
For low-angle grain boundaries, melting starts at individual
dislocations. The grain boundary can be overheated
up to a misorientation-dependent critical temperature 
at which the solid ``bridges'' between the liquid 
``pools'' break and the system becomes liquid.
Furthermore, we have found that
a hysteretic structural transition from single dislocations
to dislocation pairs can occur for intermediate values
of the misorientation. The latter, however, is generally dependent
on inclination since it is observed here for $\phi=0^\circ$ but
not $\phi=30^\circ$.

We have extracted numerically the dependence of
the disjoining potential $V(w)$ as a function of layer width $w$, and
found that its shape is qualitatively different for
high- and low-angle boundaries. For high angles, $V(w)$ 
is purely repulsive for all $w$ and reasonably well fitted by
the exponential law of Eq.~(\ref{Gw}), assumed in 
sharp-interface theories \cite{Broughton87,Rappaz03}, at least
for the largest misorientation investigated here. In contrast, 
for low-angle grain boundaries, $V(w)$ is attractive for large $w$,
but repulsive for small $w$, and exhibits a minimum that corresponds 
to the existence of a liquid layer of finite width at the melting point. 
Furthermore, this width diverges as the misorientation approaches from below 
a critical value $\theta_c$ that distinguishes these two regimes. This
divergence is smooth and reflects the progressive formation of a 
continuous premelted layer by merging of liquid-like pools and disappearance 
of solid bridges between them with increasing misorientation.

We have found that $\theta_c$ is not well
predicted by the exponential form assumed in Eq. (\ref{Gw}) with
a constant prefactor. This form
does not describe the large reduction of the grain-boundary energy
due to both the decrease of the shear modulus
at high homologous temperature, and
local melting around dislocations that is already present
for low-angle boundaries.  
We have found that, in contrast, a Read-Shockley
law for the grain-boundary energy used in conjunction
with a value
of the shear modulus at the melting point and
an effective dislocation core radius, which describes 
phenomenologically dislocation-induced melting, yields a three to
four times larger estimate of $\theta_c$ that is in better
agreement with the value obtained from PFC simulations. This estimate, 
however, is still too low due to the fact that dislocations are not 
isolated for $\theta\approx \theta_c$ as assumed in the derivation
of this law.

While this work has yielded a consistent picture of the 
thermodynamics of premelting in a microscopic model that 
can hopefully serve as a basis for developing more accurate
mesoscopic models, it has also shown that many questions
still need to be answered before a truly quantitative
description can be obtained. First, and most importantly, how does 
the disjoining potential depend generally on 
crystal structure and grain-boundary orientation characterized
by five parameters in the extension of this work to three dimensions~?
While developing a complete theoretical description of this potential 
seems difficult, there is reasonable hope that the interaction of
crystal-melt interfaces due to the overlap of density-wave
profiles for large separation ($w\gg \delta$) could be understood within the
framework of the Ginzburg-Landau theory \cite{Wuetal06,WuKar07}.
The order parameters of this theory are the complex amplitudes 
of crystal density waves in the solid and one would expect the 
range of interaction of the disjoining potential to be related 
to the rate of spatial decay of these density waves in the liquid. 
The fact that this theory can be derived from the PFC model and related 
quantitatively to experiments and MD simulations for isolated crystal-melt
interfaces \cite{WuKar07} suggests that it should provide a fruitful
theoretical framework in which to understand fundamental aspects of 
grain-boundary premelting. In particular, an asymptotic description of the 
disjoining potential for large $w$ could in principle shed light on 
the physics of the critical wetting angle.

Let us finally comment on the further perspectives of our work.
Here, we have only investigated the structural aspect of
grain-boundary premelting. It would be interesting to study
its consequences on macroscopic properties such as the
resistance to shear. In principle, shear can be incorporated
into the PFC model by modifying its equations of motion 
\cite{Steetal06}. Since the experimental evidence for 
grain-boundary premelting in pure substances is controversial, whereas 
this phenomenon is well documented in alloys, it would also be 
interesting to extend our study to this case using recently
developed PFC models for alloys \cite{Eldetal07}.

\begin{acknowledgments}
This research was supported by a CNRS travel grant as
well as by U.S. DOE through Grant No. DE-FG02-07ER46400
and the DOE Computational Materials Science Network program.
We thank Robert Spatschek for many useful discussions and
for a careful reading of the paper. 
\end{acknowledgments}

\appendix

\section{Implementation of the PFC model\label{Appendix:Numerics:Implementation}}

\subsection{Locally conserved dynamics}
The standard locally conserved dynamics for the density field $\psi$ 
is given by Eq.~(\ref{Eq:Numerics:Implementation:EOM}) as
\begin{eqnarray}
\partial_t \psi &=&(1-\epsilon)\nabla^2\psi + 2 \nabla^4 \psi + \nabla^6\psi + \nabla^2\psi^3 \nonumber\\ 
 & \equiv & \hat L \psi + f \,, \nonumber 
\end{eqnarray}
where the second equality defines the linear operator 
$\hat L \equiv (1-\epsilon) \nabla^2 + 2\nabla^4+ \nabla^6$ 
and the nonlinear function $f \equiv \nabla^2\psi^3$.

To avoid the numerically challenging gradient terms in real space, 
the equation of motion is solved in Fourier space. Multiplying
both sides of the equation by $\exp(\imath k x)$ and 
integrating over the entire volume leads to
\begin{eqnarray}
\partial_t \tilde\psi_k &=& \hat L_k \tilde \psi_k + \tilde f_k \label{Eq:Numerics:Implementation:dttildepsibar} \,, 
\end{eqnarray}
where the Fourier modes of the density are 
$\tilde \psi_k = \int  \psi \exp(\imath \vec k \vec x) dx$, 
$\hat L_k = (\epsilon-1) k^2 + 2 k^4 - k^6$ is the linear operator 
in Fourier space, and $\tilde f_k$ is the Fourier 
transform of the nonlinear function $f$.

Furthermore, an implicit integration scheme is used which allows us
to use larger time steps. Instead of solving 
Eq.~(\ref{Eq:Numerics:Implementation:dttildepsibar}) directly, 
it can be rewritten by using the ansatz 
$\tilde \psi_k = u(t) \exp(\hat L_k t)$. One then obtains
\begin{eqnarray}
\partial_t \tilde\psi_k &=& \hat L_k \exp(\hat L_k t) u(t) + (\partial_t u) \exp(\hat L_k t) 
\nonumber \\
& = & \hat L_k \exp(\hat L_k t) u(t) + \tilde f_k \nonumber \,,
\end{eqnarray}
so that $\partial_t u(t)  = \exp(-\hat L_k t) \tilde f_k$. Integrating over time from $t$ to $t+\Delta t$ gives
\begin{eqnarray}
u(t+\Delta t) - u(t) & =& \int_t^{t+\Delta t} dt' \exp(-\hat L_k t') \tilde f_k(t') \nonumber 
\end{eqnarray}
and with $u(t) = \exp(-\hat L_k t) \tilde\psi_k(t)$ in terms of $\tilde\psi_k$
\begin{eqnarray}
&\exp[-\hat L_k (t+\Delta t)] \tilde\psi_k(t+\Delta t) - \exp(-\hat L_k t) \tilde\psi_k(t) &\nonumber \\  
&=&\nonumber \\
& \int_t^{t+\Delta t} dt' \exp(-\hat L_k t') \tilde f_k(t')  \,\,\,. & \nonumber 
\end{eqnarray}
Even if $\tilde f_k$ is not known as a function of $t$, it can be expanded in a good approximation around $t'=t$, leading to
\begin{eqnarray}
\tilde \psi_k(t+\Delta t) &=& e^{\Delta t\hat L_k} \tilde \psi_k(t) \nonumber \\
&+& \,\, e^{\hat L_k(t+\Delta t)}\int_t^{t+\Delta t} dt' e^{-\hat L_k t'}
\nonumber \\ 
&\times& \, \left[\tilde f_k(t) + \frac {\tilde f_k(t) - \tilde f_k(t-\Delta t)}{\Delta t} (t' -t)\right] \nonumber \\
&=& e^{\Delta t\hat L_k}  \tilde \psi_k(t) + \frac {\tilde f_k(t)}{\hat L_k} \left(e^{\hat L_k \Delta t} -1 \right)\nonumber \\
&+& \frac {\tilde f_k(t)-\tilde f_k(t-\Delta t)}{\Delta t \hat L_k^2} \left(e^{\hat L_k \Delta t}-1 -\Delta t \hat L_k\right)  \,.
\nonumber\\ \label{Pfc_Boston_Eq_EOM-in-Code}
\end{eqnarray}

\subsection{Non-local globally conserved dynamics}
To accelerate the search for the equilibrium solution, a different 
non-local dynamical formulation can be used where the dynamics
depends globally on the density field, as opposed to locally in
the standard conserved dynamics. The global conservation of the 
order parameter has then to be ensured by a Lagrange multiplier. 

The conservation condition for $\psi$ is given as
\begin{eqnarray}
\int \psi(\vec x) d\vec x - L_xL_y\bar\psi  & = & 0 \nonumber \,,
\end{eqnarray}
where $\bar\psi=1/(L_xL_y)\int \psi(\vec x)$ is the average density.
The free energy, including the constraint, can then be written as
\begin{eqnarray}
\widetilde{\mathcal{F}} & = & \mathcal{F} + \mu \left[\int \psi(\vec x) d\vec x - L_xL_y\bar\psi \right] \nonumber \,.
\end{eqnarray}
where $\mu$ is the Lagrange multiplier.

The equation of motion becomes
\begin{eqnarray}
\partial_t \psi &=& - \frac{\delta \mathcal{F}}{\delta \psi} + \mu  \nonumber \\
&=&\left[(\epsilon-1) - 2 \nabla^2 - \nabla^4\right]\psi - \psi^3 + \mu \nonumber\\
 & \equiv & \hat L \psi + f \,, \nonumber 
\end{eqnarray}
where now $\hat L  \equiv (\epsilon-1) - 2 \nabla^2 - \nabla^4$ 
and $f \equiv  - \psi^3 + \mu $. In Fourier space, the linear operator and 
the nonlinear function are given as  $\hat L_k  = (\epsilon-1) + 2 k^2 - k^4$ 
and $\tilde{f}_k = -\tilde{\psi}^3_k + \tilde{\mu}_k$, where 
$\tilde{\psi}^3_k$ is the Fourier transform of $\psi^3$ and $\tilde{\mu}_k$ 
is that of $\mu$. Since $\mu$ is a  constant, 
$\tilde \mu_k \propto \delta(k)\mu$.  With this $\hat L_k$ and 
$\tilde{\mu}_k$, the implicit integration scheme as given in 
Eq.~(\ref{Pfc_Boston_Eq_EOM-in-Code}) can be used.

The Lagrange multiplier $\mu$ can be obtained from the condition
\begin{eqnarray}
 0 = \partial_t \bar\psi & = &
    \frac {1}{L_xL_y} \int \partial_t \psi(\vec x) d\vec x \nonumber\\ 
   & = & 
    \frac {1}{L_xL_y} \int \left[- \frac{\delta \mathcal{F}}{\delta \psi} + \mu \right] d\vec x  \nonumber \,,
\end{eqnarray}
or
\begin{eqnarray}
 {\mu} &=& \frac {1}{L_xL_y} 
    \int \frac {\delta \mathcal{F}}{\delta \psi}  d\vec x \nonumber \\
       &=& \frac {1}{L_xL_y}
    \int \left[ (1-\epsilon)\psi(\vec x) + \psi^3(\vec x)\right] d\vec x  \nonumber \,,
\end{eqnarray}
since the integral over the gradients is zero for a periodic system.

\section{Boundary conditions}
The two-dimensional hexagonal periodic solution given by 
Eq.~(\ref{Pfc_Boston_Eq_SolHex}) exhibits close-packed rows
of density peaks along the $x$ direction and can be described
by two basis vectors
$\vec a = a \left(1,0\right)$ and $\vec b = a\left(1/2,\sqrt{3}/2\right)$,
where $a=2\pi/q=4\pi/\sqrt{3}$ is the ``lattice spacing'' (the
spacing between density peaks). When the entire structure is
rotated by an angle $\Theta$ 
($x \to x\cos\Theta + y\sin\Theta$ and $y \to -x \sin\Theta + y \cos\Theta$), 
the rotated basis vectors are
\begin{subequations}
\begin{eqnarray}
\vec a &=& a \left(\begin{array}{c} \cos\Theta \\ - \sin\Theta \end{array}\right) \,\quad \mathrm{and}  \\
\vec b &=& a \left(\begin{array}{c} \frac 12 \cos\Theta + \frac{\sqrt{3}}{2}\sin\Theta \\ 
                                    -\frac 12 \sin\Theta + \frac{\sqrt{3}}{2}\cos\Theta
\end{array}\right) \,.
\end{eqnarray}
\end{subequations}
In order to exactly fit a periodic structure into the simulation
box, a displacement of once the box size along the box axes must
correspond to an integer number of steps along the two basis
vectors, that is, we must have
\begin{subequations}
\begin{eqnarray}
n \vec a - m\vec b &=& \left(\begin{array}{c} L_x \\ 0 \end{array}\right)\quad \mathrm{and} \\
- i \vec a + j\vec b &=& \left(\begin{array}{c} 0 \\ L_y \end{array}\right)\,,
\end{eqnarray}
\end{subequations}
where (without loss of generality) $0<\Theta<\pi/3$, and $n$, $m$, 
$i$ and $j$ are integer numbers. The minus signs have been chosen
by convention such that the conditions can be satisfied with positive
integers.

From the components of the above vector equations that have 
a zero on the right hand side, we obtain two conditions for 
the angle,
\begin{subequations}
\begin{eqnarray}
\tan \Theta &=& \sqrt{3} \frac {m}{2n+m} \quad \mathrm{and}\\
\tan \Theta &=& \frac{1} {\sqrt{3}}\frac{ 2i-j }{j} \,.
\end{eqnarray}
\end{subequations}
Only angles can be simulated for which four suitable integers
can be found that satisfy both conditions. Note that with
sufficiently large integers, any angle can be approximated
to arbitrary precision. Once the four integers are determined,
the dimensions of the simulation box are given by
\begin{subequations}
\begin{eqnarray}
L_x &=& a \left[\left(n-\frac{m}{2}\right) \cos\Theta - m\frac{\sqrt{3}}{2} \sin \Theta \right], \label{eq:Xlength} \\
L_y &=& a \left[ \left(\frac j 2 - i \right)\sin\Theta + j\frac{\sqrt{3}}{2}\cos\Theta \right]. \label{eq:Ylength}
\end{eqnarray}
\end{subequations}
For the numerical treatment, the equations have to be discretized.
In order to accommodate both conditions for the system size, in
general slightly different grid spacings have to be used along 
the $x$ and $y$ directions since the ratio $L_x/L_y$ may be
irrational and cannot be well approximated by a single grid spacing.
We always chose grid spacings $\Delta x$ and $\Delta y$ that are
close to $\pi/4$, as in previous PFC studies \cite{Eldetal02,Eldetal04}.
We have checked by repeating selected runs with different choices for
the discretization that grid effects are negligible for all the 
results presented here.

In the presence of grain boundaries, the density field is 
no longer periodic in the $y$ direction, and
Eq.~(\ref{eq:Ylength}) for $L_y$ does not apply. In this case,
no simple condition for $L_y$ can be given that ensures a
strain-free bulk solid, since this would require a detailed
knowledge of the grain-boundary structure. However, this
condition is not as stringent as for single crystals since
there is an additional degree of freedom: the dislocations
present at the grain boundaries can move along the boundaries 
in response to bulk stress until a minimum of the energy
is reached, which implies a relaxation of the bulk stress.
Even if there is only a finite number of dislocation positions
that correspond to a local energy minimum (this number scales 
as $d/a$, where $d$ is the distance between dislocations), 
as long as $L_y$ is chosen to be much larger than $L_x$, the residual 
bulk stresses should be very weak. Indeed, we have varied $L_y$ 
by small amounts for several sets of parameters, and never 
found significant variations in the free-energy density.

It should be noted that the condition on $L_x$, Eq.~(\ref{eq:Xlength}) 
still applies. For low-angle grain boundaries, the numbers $m$
and $n$ can easily be related to explicit dislocation models. 
For instance, consider the $\phi=0^\circ$ inclination: 
the number $m$ corresponds to the number of
close-packed planes that originate at the grain boundary
for each of the two tilted grains; the total number of
edge dislocation is therefore equal to $2m$. In turn, $n$
indicates the number of steps that have to be taken along
a close-packed row before a site that is geometrically
equivalent to the starting site can be reached by $m$
steps along the basis vector $\vec b$. While the average 
spacing $d$ between dislocations is therefore always equal 
to $L_x/(2m)$, the minimum-energy configuration does not
always correspond to equal spacings between dislocations.
For instance, for $m=1$, $n$ even yields two dislocations
that are evenly spaced, whereas $n$ odd corresponds to a
grain boundary where a slightly larger and smaller spacing
alternate along the interface. Such configurations are
well known \cite{ReadShockley} and constitute a local energy
minimum. We did not notice any considerable difference between
the behaviors of these two types of grain boundaries. This
is to be expected since, due to the condition on $L_x$
in Eq.~(\ref{eq:Ylength}), the system is still globally 
strain-free far from the grain boundary.


\begin{thebibliography}{99}
\bibitem{remark1} In the $(N,p,T)$ ensemble, the volume is not
constant; therefore, strictly speaking, Gibbs free energies per 
particle should be used rather than per unit volume. However,
it can be shown that the two approaches are equivalent here.
For simplicity, we neglect the interfacial energies and write
the Gibbs free energy as $G=N_s g_s + N_l g_l$, where $N_s$
and $N_l$ are the numbers of particles in the solid and the
liquid (with $N_l+N_s=N$ fixed), and $g_s$ and $g_l$ are the
free energies per particle in the solid and the liquid.
Subtracting the value for a homogeneous solid, $Ng_s$, and
dividing by the total surface area $A$ of the liquid film,
we obtain $G_{\mathrm{exc}} = (N_l/A)(g_l-g_s)$ which can be rewritten
as $G_{\mathrm{exc}} = (N_lV_l/A)[(g_l-g_s)/V_l]$ where $V_l$ is the
volume per particle in the liquid. The expression used in 
Eq.~(\ref{Gw}) is obtained by identifying $w=N_lV_l/A$ and
$\Delta G=(g_l-g_s)/V_l$.
\bibitem{remark2} The analogous quantity in the $(N,p,T)$ ensemble 
for fixed pressure and varying temperature is $u=(T-T_m)/(L/c_p)$, 
where $L$ is the latent heat of fusion and $c_p$ the specific heat 
at constant pressure.
\bibitem{remark3} We have $G=Y/[2(1+\sigma)]$, where $G$, $Y$, and
$\sigma$ are the shear modulus, bulk modulus, and Poisson's ratio
in three dimensions, respectively. In two dimensions, we have 
$\sigma_2 = \sigma/(1-\sigma)$ and $Y_2 = Y/(1-\sigma^2)$.
Inserting these relations in Eq.~(\protect\ref{Eq:Pfc:RS2}) transforms 
the prefactor into $aY_2/(8\pi)$, which is identical to the
result of Ref.~\protect\cite{Eldetal04}, taking into account that
the modulus of the Burgers vector of a standard edge dislocation 
in the two-dimensional hexagonal structure is equal to the lattice 
spacing.
\end{thebibliography}
\end{document}